\documentclass[1p]{elsarticle}
\usepackage{graphicx}
\usepackage{array}
\usepackage[table]{xcolor}
\usepackage{multirow}
\usepackage{amssymb}
\usepackage{booktabs}
\usepackage{makecell}
\usepackage{empheq}
\usepackage{natbib}
\usepackage{appendix}
\usepackage{subfig}
\usepackage{xcolor}
\usepackage{eurosym}
\usepackage{url}
\usepackage{float}
\usepackage[T1]{fontenc}
\usepackage[utf8]{inputenc}
\usepackage{lmodern}
\usepackage{pdflscape}

\usepackage{enumitem}
\usepackage{nomencl}
\makenomenclature

\usepackage{mathtools}

\newdefinition{rmk}{Remark}
\newtheorem{proof}{Proof}
\newtheorem{ass}{Assumption}
\newtheorem{prop}{Proposition}

\usepackage{hyperref}
\newcommand{\sref}[2]{\hyperref[#2]{#1 \ref*{#2}}}

\let\oldref\ref
\renewcommand{\ref}[1]{(\oldref{#1})}
\usepackage[nameinlink, capitalise, noabbrev]{cleveref}
\hypersetup{
	colorlinks,
	citecolor=blue,
	filecolor=blue,
	linkcolor=blue,
	urlcolor=blue}
\crefname{prop}{Proposition}{Propositions}
\crefname{thm}{Theorem}{Theorems}
\crefname{cor}{Corollary}{Corollaries}
\crefname{ass}{Assumption}{Assumptions}
\makenomenclature
\usepackage{etoolbox}
\renewcommand\nomgroup[1]{%
	\item[\bfseries
	\ifstrequal{#1}{A}{Sets}{%
		\ifstrequal{#1}{F}{Futures market parameters}{%
			\ifstrequal{#1}{V}{Variables}{%
				\ifstrequal{#1}{M}{Spot market and RES parameters}{}}}}%
	]}
\makeatletter
\def\bs{\expandafter\@gobble\string\\}
\def\lb{\expandafter\@gobble\string\{}
\def\rb{\expandafter\@gobble\string\}}
\def\@pdfauthor{C.V.Radhakrishnan}
\def\@pdftitle{elsarticle.cls -- A documentation}
\def\@pdfsubject{Document formatting with elsarticle.cls}
\def\@pdfkeywords{LaTeX, Elsevier Ltd, document class}

\DeclareRobustCommand{\LaTeX}{L\kern-.26em%
	{\sbox\z@ T%
		\vbox to\ht\z@{\hbox{\check@mathfonts
				\fontsize\sf@size\z@
				\math@fontsfalse\selectfont
				A\,}%
			\vss}%
	}%
	\kern-.15em%
	\TeX}
\journal{OMEGA (The International Journal of Management Science)}

\begin{document}

\begin{frontmatter}

\title{Contract design in electricity markets with high penetration of renewables: A two-stage approach}

\author[1,2]{Arega Getaneh Abate\corref{cor1}}
\ead{a.abate004@unibs.it}
\author[1]{Rossana Riccardi}
\ead{rossana.riccardi@unibs.it}
\author[3,4]{Carlos Ruiz}
\ead{caruizm@est-econ.uc3m.es}
\cortext[cor1]{Corresponding author}
\address[1]{Department of Economics and Management,
University of Brescia, 74/B Brescia, Italy}
\address[2]{Scuola Normale Superiore di Pisa, Piazza del Cavalieri, 7-56126 Pisa Italy}
\address[3]{Department of Statistics, University Carlos III de Madrid, Avda. de la Universidad 30, 28911-Leganés, Spain.}
\address[4]{UC3M-BS Institute for Financial Big Data (FiBiD), Universidad Carlos III de Madrid, 28903, Getafe, Madrid, Spain.}
\begin{abstract}
The interplay between risk aversion and financial derivatives has received increasing attention since the advent of electricity market liberalization. One important challenge in this context is how to develop economically efficient and cost-effective models to integrate renewable energy sources (RES) in the electricity market, which constitutes a relatively new and exciting field of research. This paper proposes a game-theoretical equilibrium model that characterizes the interactions between oligopolistic generators in a two-stage electricity market under the presence of high RES penetration. Given conventional generators with generation cost uncertainty and renewable generators with intermittent and stochastic capacity, we consider a single futures contract market that is cleared prior to a spot market where the energy delivery takes place. We introduce physical and financial contracts to evaluate their performance and assess their impact on the electricity market outcomes and examine how these depend on the level of RES penetration. Since market participants are usually risk averse, a coherent risk measure is introduced to deal with both risk neutral and risk averse generators. We derive analytical relationships between contracts, study the implications of uncertainties, test the performance of the proposed equilibrium model and its main properties through numerical examples. Our results show that overall electricity prices, generation costs, profits, and quantities for conventional generators decrease, whereas quantities and profits for RES generators increase with RES penetration. Hence, both physical and financial contracts efficiently mitigate the impact of uncertainties and help the integration of RES into the electricity system. However, each type of contract has different quantitative impacts on the market outcomes. Moreover, risk aversion increases profit and decreases generators' risk under different levels of competition by shielding them from the risk associated with cost, demand, and RES uncertainties.
\end{abstract}  
\begin{keyword}
	Contract designs; Contracts for differences;  Game-theoretical model;  Renewable penetration;  Risk aversion
\end{keyword}

\end{frontmatter}
 
\section{Introduction}
\label{P_11}
The increasing concern about climate change, limited electricity generation sources, reliability of electricity supply, and security of electricity service while integrating renewable energy sources (RES) into the electricity system have been extensively important topics in energy economics. Such models have emerged after the liberalization of the electricity sector in several countries since the early 1990s. Liberalization has not only brought the active role of demand and supply in the electricity market but also has increased inter-regional trades so that greater competition in electricity generation has spurred efficient production, distribution, and transmission of electricity \citep{eydeland2003energy,international2005learning,wang2018two}. 
In recent years, the transition from fossil-based to RES and sustainable energy is favored by international agreements and national plans with favorable economics, ubiquitous resources, scalable technology, and significant socio-economic benefits \citep{jurasz2017integrating,gielen2019role}. However, the transition and integration of RES in the electricity system further complicate the electricity system management to maintain supply-demand balance at all times. This is because the stochasticity and intermittent natures of RES lead to unstable supply and price volatility which, in turn, make market participants bear increased financial risks \cite{alshehri2021centralized, woo2011impact}. In addition, large RES penetration may have a policy paradox in a liberalized market design due to its zero marginal cost and non-dispatchability which may result in depressed and more volatile electricity prices, which in turn makes renewables incentives to be more expensive and counterproductive \citep{blazquez2018renewable,imran2014technical}. In other words, market participants could be exposed to more risk at all levels of the supply chain (production, transmission, and distribution), which further complicates risk management \citep{bruno2016risk,falbo2019optimal}.  Empirical literature in the electricity market has provided tools and models to deal with the distinctive electricity features and risk management through derivatives, which are common in commodity markets \citep{bushnell2007oligopoly,carrion2007forward,anderson2008forward,baringo2013risk,philpott2016equilibrium}. For instance, \citet{philpott2016equilibrium} show how a risk averse social plan is equivalent to competitive equilibrium when agents use dynamic coherent risk measures and there are enough market instruments to hedge inflow uncertainty.
   
 For electricity generators, the derivatives markets work by allowing them to commit part of their production through forward or/and bilateral contracts to reduce the impact of major uncertainties on electricity outcomes. Therefore, generators, retailers, and consumers can agree on prices, quantities, delivery times, and other conditions within the framework of the contracts. There is a vast number of different contracts, from plain forwards to interruptible contracts which can improve competition in the spot market by applying sophisticated risk management modeling \citep{willems2010market,bushnell2007oligopoly}.

In particular, futures contracts become more liquid and relevant in electricity trading to hedge price risks given the non-storability of electrical energy and its inelastic demand. Futures markets allow trading products spanning a large time horizon, such as one month or a quarter.
Since the electricity market depends on key variables such as fuel prices, electricity demand, and weather, introducing financial contracts in generators' profit function with different generation technologies (conventional and RES) under uncertainty is not an easy task.  Moreover, as the stability of power grids (frequently off equilibrium) is always an issue with the high penetration of RES, designing electricity markets considering the dynamic nature of renewables, which require a reserve capacity based on controllable sources) and financial contracts liquidity (forward/futures) is a nontrivial task \cite{gouveia2014effects}.

From a methodological perspective, different mathematical models have been proposed for determining optimal strategies in the electricity market with RES penetration using financial contracts. Some of these consider deterministic models, and others deal with uncertainty by using stochastic programming or robust optimization  approaches either considering market power and/or with Nash competitive games \citep{allaz1992oligopoly, allaz1993cournot,botterud2011wind,moller2011balancing,bruno2016risk}.
After the pioneering work of \citet{allaz1992oligopoly} that deals with oligopolistic generation traded on a forward market for risk hedging, extensions with uncertainty and with different modeling approaches have been introduced \citep{allaz1993cournot,willems2010market, bushnell2007oligopoly,jurasz2017integrating}.
  
However, there are still important gaps in the literature: (1) in solving the analytical relationship between futures and spot markets contracts by considering RES penetration and risk aversion simultaneously, (2) in understanding the challenges entailed with large RES integration with better modeling techniques and a coherent risk management, and (3) in considering generators’ behavior with risk hedging via physical and financial assets. In this paper, we contribute to fill these gaps by investigating how the risk associated with the interactions between RES and conventional generators could be managed, how participants risk aversion could affect electricity market outcomes, and how financial derivatives (both physical and financial) can hedge generators’ risk exposures under different market configuration under uncertainty. More specifically, the main motivations are: (1) to better understand the impact of high renewables penetration on existing liberalized market designs where the market operates on the assumption that electricity generation has a range of positive marginal costs. This further allows us to examine sustainability adaptation goals (increasing RES penetration) while addressing an increasing power demand, and 
(2) to explore explicitly how this renewable penetration interacts with the relationship between spot and futures markets under different levels of competition, generators’ risk aversion, and different contract designs. This gives us insights on how market participants and risk managers can efficiently approach a market under uncertainty by using financial contracts. 

To this end, we propose a two-stage game-theoretical model where each oligopolistic generator optimizes its profit by integrating the futures (physical and financial) and spot markets while conjecturing the impact its production may have on other generators' quantities and market prices. Uncertainties are managed with contracts and a coherent risk measure is introduced in the generators' decision processes. The proposed equilibrium model characterizes the interactions between generators with different levels of market power (à la Cournot, and perfect competition) in a two-stage electricity market. We consider a single futures market that is cleared before a spot market by introducing both physical and financial contracts to evaluate their impact and performance in the equilibrium market outcomes with high RES penetration.

First, we develop a general model (GM) with a futures market that requires physical delivery when the spot market takes place. Then, we consider a market where the futures market makes use of contracts for differences (CFD). 
  CFD is a financial instrument that makes it possible for the players in the market to hedge against the difference between the futures price and the spot market price in a future time period, and serves as a market coordination mechanism.  
For comparison we also analyze the spot market without the presence of futures trading. To deal with both risk averse and risk neutral generators, we introduce a coherent risk measure (Conditional Value at Risk, CVaR) which has been extensively used in the literature after its introduction by \citet{rockafellar2000optimization} for portfolio optimization problems\footnote{Although VaR is a well-known risk measure in economic problems, it is a non-coherent risk measure suffering from non-convexity, non-smoothness, subadditivity, etc., which makes it undesirable in optimization programs. To avoid this problem, there is an attractive alternative risk measure identified as CVaR also known as average value at risk or mean shortfall.  CVaR is a risk assessment technique often used to reduce the probability a portfolio will incur large losses. This is performed by assessing the likelihood (at a specific confidence level $\alpha$) that a specific loss will exceed the value at risk. That is, CVaR is derived by taking a weighted average between the VaR and losses exceeding the VaR.}. We derive analytical relationships between contracts and study the implications of demand, production cost, and RES capacity uncertainties on electricity market outcomes. Finally, the performance of the proposed equilibrium model and its main properties are tested with extensive numerical simulations. Our results show that overall electricity prices, generation costs, profits, and quantities decrease for conventional generators, whereas quantities and profits increase for RES generators with respect to the level of RES penetration. Despite physical and financial contracts exhibit different impacts on the resulting equilibrium market outcomes, they both efficiently mitigate the impact of uncertainties and facilitate the integration of RES into the electricity system. Moreover, risk aversion increases profit and decreases generators' risk in the general model for both the Cournot and perfect competitions, which is a counter-intuitive result. That is because contracts increase market participants' willingness to participate in the market  as it shields them from the risk associated with production cost, demand, and RES uncertainties. 

The paper is organized as follows. \Cref{L_1200} reviews the current state-of-the-art in the electricity market equilibrium models with renewables and underlines our contribution. \Cref{L_1201} formulates the model for the two-stage problem, showing how it can be reformulated from stage-two to stage-one in a stochastic programming framework. \Cref{L_1203} shows how the parameters are generated, discusses the simulation results, and comments on the numerical results, and their implications. Finally, \Cref{L_1202} concludes with some relevant remarks.
\section{Literature review and contributions}\label{L_1200}
\subsection{Literature review}
In the post-liberalized electricity market, the economics and regulation of electrical energy systems that include an integral amount of stochastic renewable generation have been studied via mathematical models using financial derivatives under uncertainty \citep{collins2002economics,hasan2008electricity,kettunen2009optimization, martin2014stochastic, duenas2014gas, morales2014clearing}. These financial instruments are traded in over the counter (OTC) and/or by exchange markets as energy commodities, and guarantee the delivery of an established amount of electricity over a specific future period, with physical delivery or financial settlement. The literature on electricity market equilibrium modeling with financial derivatives has rapidly developed because of the growing need for obtaining models that describe electricity market behavior accurately and the need to integrate intermittent RES in the electricity system.   Overall, three streams of research are closely related to this study: \textit{futures and spot markets for risk hedging in the electricity market, high renewables penetration into the electricity systems}, and \textit{stochastic models in the electricity market}. In what follows, we review studies relevant to each stream and highlight the differences between this study and the existing literature.
 
\subsubsection{Futures and spot markets in electricity}
The futures market has become relevant for trading electricity as it helps to hedge the volatility of spot market prices based on prior agreements. The futures contracts have the spot price as an underlying reference in both physical and financial settlements, where generators can reduce their risk exposure for later delivery \citep{allaz1992oligopoly,allaz1993cournot,kettunen2009optimization, OMIE2020,EEX2020}.  Futures contracts assure a fixed price of electricity in the future while spot market contracts are subject to uncertainties. 

\citet{allaz1992oligopoly}, with a simple two-period classical model of an oligopoly producing a homogeneous good, shows forward contracting can be an effective tool in the hands of noncompetitive producers. \citet{allaz1993cournot} develop a model with two Cournot duopolies and forward market contracts. They show that at equilibrium, each Cournot player will sell forward which may worsen their position if they do not fully understand when and how to use these contracts prior to their production. \citet{powell1993trading} develops a model of contracting in the electricity industry where the generators are price setters in the futures market, and quantity setters in the spot market with risk aversion. He concludes that the futures price is higher than the expected spot market price. 

\citet{mendelson2007strategic} analyze forward and spot trading under uncertainties among supply chain participants by utilizing new demand and cost information. \citet{bushnell2007oligopoly} extends the Allaz and Vila model to multiple firms with increasing costs to demonstrate that suppliers’ market power plays a key role in the interaction between forward and spot trading. \citet{de2013liquidity} analyze the interaction between financial instruments with spot electricity markets with a generalized Nash equilibrium framework by accounting for payoffs and an incomplete risk market endogenously. 

There are other examples where financial derivatives have been used extensively for risk management in the electricity market \citep{le2002forward,willems2010market,su2007analysis,anderson2008forward,ruiz2012equilibria,de2013liquidity,martin2014stochastic,ralph2015risk}. These contracts have been used to achieve different objectives from different modeling perspectives: for instance, physical and financial contracts with different levels of competition in electricity market \citep{willems2010market}, futures trading with retailers to achieve social welfare with market power \citep{anderson2008forward}, electricity markets without price volatility \citep{su2007analysis}, market equilibrium model with asymmetric producers \citep{le2002forward}, the equilibrium in futures and spot markets with oligopolistic generators and conjectural variations \citep{ruiz2012equilibria}, effects of futures markets on the investment decisions of a strategic electricity producer \citep{martin2014stochastic}, a risky design game (a stochastic Nash game) problem in a complete risk market using financial products \citep{ralph2015risk}, are some relevant contributions. However, none of these have solved the analytical relationship between futures and spot markets contracts combining high renewables penetration and risk aversion simultaneously. Even when they analyze contract designs in detail, either they do not consider risk using a coherent measure or do not address the generation mix. Thus, this paper tackles these gaps by deriving the relationship between futures and spot markets contracts, settled physically and financially and with different levels of competition under uncertainty.  In doing so, it accounts for the advocate of sustainable development by considering high RES penetration into the generation mix where financial contracts protect generators from the embedded risks of both generation and demand.

\subsubsection{High RES penetration}
Because of the growing concern on conventional generation's impact on the environment and its high generation cost, the electricity generation mix has been significantly changing with strict policies towards sustainable energy adaptation. In particular, the increasing penetration of RES in electricity systems is leading towards a green economy and lowering electricity prices. Nevertheless, due to high uncertainty on RES capacity, zero marginal cost and non-dispatchability\footnote{The assumption with renewables zero marginal cost, mainly with wind and solar, is that once they start producing, the marginal cost to continue operating is almost zero. This contradicts the idea of liberalization that assumes a positive marginal production cost, and dispatchability for the market to work properly.},  integrating RES into the electricity systems requires a risk averse dispatch of resources to account for the stochastic availability of renewable capacity and to hedge the price volatility \citep{zhang2015distributed,zou2017electricity}.

In relation to this, governments have been creating better conditions (with specific contracts, green certificates, feed-in-tariffs, and other market-based approaches) towards investing in renewables to be integrated in conventional electricity systems \citep{street2009risk,boomsma2012renewable, verbruggen2012assessing,mitridati2021design}. 

The electricity market with high RES penetration has been studied with different approaches. For instance, with a two-stage energy management strategy under uncertainty \citep{wang2018two}, with multistage stochastic programming by combining real options and forward contracts for risk management in renewable investment \citep{bruno2016risk}, with a risk-based energy management of renewable-based microgrids using information gap decision theory \citep{mehdizadeh2018risk}, are some of the contributions. Since risk and uncertainty are the main challenges considering RES, \cite{nguyen2014risk} design a CVaR model that enables the microgrid aggregator to exploit the flexibility of the load to mitigate the negative impacts of the uncertainties due to RES generation and electricity prices.  
Hence, CVaR is a well-known risk measure that has been widely used in various energy management problems for different entities in the electricity market such as for retailers \citep{carrion2007forward,garcia2013sale}, for producers \citep{conejo2008optimal,morales2010short,baringo2013risk,abate2021contracts}, for distributors \citep{safdarian2013stochastic}, for optimal power flow \citep{zhang2013robust} and for coordinated energy trading problems \citep{al2011coordinated,de2013optimal}.  This paper departs from the literature by tackling uncertainties with contracts, so that electricity market participants act upon the received information, and understand the challenges entailed with large RES integration with better modeling techniques and a coherent risk management.
 
\subsubsection{Stochastic equilibrium models in the electricity market}

Equilibrium models in electricity market can be applied to represent the overall market behavior taking into consideration the competition among all participants, such as, Nash equilibrium (Cournot and supply function equilibria). In relation to this, \citet{ventosa2005electricity} discuss research developments in electricity equilibrium models based on: degree of competition, time scope of the model, uncertainty modeling, inter-period links, transmission constraints, generating system representation and market modeling.

Specifically, in the optimization research community, various equilibrium models in the electricity market and related problems under uncertainty have drawn increasing attention, particularly on short-term operations, capacity expansion, optimal dispatch, optimal power flow, RES integration, risk management, and decarbonization, which are very important topics needed to analyze electricity market designs. Novel mathematical model formulations and numerical methods have been proposed to deal with those problems. \citep{gurkan1999sample,vehvilainen2005stochastic,galiana2010emission,lin2010stochastic,pozo2011finding,aid2011hedging,zhang2013robust,schroder2014electricity,djeumou2019applications,dvorkin2019chance,hao2020does,mitridati2021design}. For example, \citet{gurkan1999sample} explore the application of sample-path to the investments in gas production by formulating a stochastic variational inequality problem, and \citet{vehvilainen2005stochastic} present a stochastic factor-based approach to mid-term modeling of spot market prices in deregulated electricity markets. \citet{galiana2010emission} propose a model on emissions trading that can be viewed as a mathematical program subject to a Nash equilibrium problem, which in turn, is subject to the Cournot-Nash equilibrium conditions of an hourly oligopolistic electricity market. \citet{aid2011hedging} use an equilibrium model to study the relationship between forward, spot and retail markets, and conclude both vertical integration and forward hedging retail prices decrease under demand uncertainty. Finally, \citet{dvorkin2019chance} develop a chance-constrained stochastic market design that enables to produce a robust competitive equilibrium and internalize uncertainty of the RES in the price formation process.
However, modeling uncertainty in the electricity market by integrating conventional and RES oligopolistic generators, considering generators with risk hedging via physical and financial assets is still an important challenge.  This inspired us to develop game-theoretical stochastic models where the performance of different contracts is analyzed in a two-stage electricity market under uncertainty. The models are analytically and numerically solved while managing uncertainties with a coherent risk measure and financial contracts. Moreover, conventional and RES generators are considered, and equilibrium market outcomes are compared for different levels of competition with respect to RES penetration.

\subsection{Contribution of the paper}

In this paper, we propose three game-theoretical equilibrium models based on: a futures contract with physical delivery, a contract for differences with financial settlement, and a spot market contract with no futures. The GM and CFD are solved based on a pre-existing futures market contract, assuming each generator has an estimation (conjecture) of the impact that its decision may have on the other generators' strategies. The spot market with no futures is derived for comparison purposes.  Most importantly, the three spot market equilibria are solved in closed form. The analytical relationship between futures and spot contracts, and implications of uncertainties on market outcomes, are examined from risk aversion, generation technologies, and levels of competition perspectives. The models are tested with multiple case studies based on calibrated data from the Spanish electricity market, and the comparison of physical and financial contracts is made focusing on systems with high RES penetration. Therefore, the main contributions of the paper are summarized as follows:
\begin{enumerate}
\item [(1)] Modeling conventional and RES generators’ profit-maximizing problems with a pre-existing financial contract, and risk aversion using the CVaR. Conjectural variations and Nash-equilibrium approaches are assumed so that all generators optimize their profit simultaneously. We consider two strands of contracts: futures with physical delivery and contracts for differences settled financially. 
\item [(2)] Proposing game-theoretical models where each generator solves a two-stage stochastic problem and examining generators’ exposure to uncertainties in demand, conventional generators’ production cost, and RES generation availability. The impact of these uncertainties on electricity market outcomes are examined for both sets of generators with respect to high RES penetration. 
\item[(3)] Deriving the spot market equilibrium in closed form to deal with the futures market where the posterior numerical solution of the overall equilibrium is reformulated as a nonlinear problem (NLP) by applying the method in \citet{leyffer2010solving}, and the CVaR.
\item [(4)] Illustrating the analytical results using numerical examples that analyze the quantitative and qualitative impacts of RES penetration, levels of competition, and players’ risk aversion on market outcomes.   
\end{enumerate}

\section{Problem formulation}\label{L_1201}
We consider the profit optimization of two types of electricity generators (conventional and RES, which includes solar, wind, or both) that trade their electricity in a two-stage electricity market. In the first stage, generators sell their electricity in a futures market. In the second stage, they participate in a spot market where they trade the remaining electricity.
 We derive three models.  The first model is the general model (GM) with physical delivery, and the second model is based on contracts for differences (CFD)-settled financially. Finally, for comparison, we compute a third equilibrium model without the presence of the futures market (only spot market). 
 
	\nomenclature[A]{$I$}{Set of conventional generators ranging from $i =1,..., |I| $}
	\nomenclature[A]{$J$}{Set of renewable resource generators ranging from $j=1,...,|J|$}
	\nomenclature[A$W$]{$\Omega$}{Set of scenarios ranging from $\omega =1,...,|\Omega|$}
	\nomenclature[F]{$\beta^F$}{Electricity price demand slope in the futures market}
	\nomenclature[M]{$\beta_{\omega}^{S}$}{Electricity price demand slope in the spot market}
	\nomenclature[M]{$\gamma_{\omega}^S$}{Electricity price demand intercept in the spot market}
	\nomenclature[M]{$Q_{j\omega}$}{RES total production for generator $j$ [MWh]}
	\nomenclature[F]{$\gamma^{F}$}{Electricity price demand intercept in the futures market}
	\nomenclature[V]{$P^F$}{Electricity price [\euro/MWh] in the futures market}
	\nomenclature[V]{$P_{\omega}^S$}{Electricity price [\euro/MWh] in the spot market}
	\nomenclature[V]{$q_i^F$}{Futures market electricity quantity [MWh], conventional generator}
	\nomenclature[V]{$q_{i\omega}^S$}{Spot market electricity quantity [MWh], conventional generator}
		\nomenclature[V]{$q_j^F$}{Futures market electricity quantity [MWh], RES generator}
	\nomenclature[V]{$q_{j\omega}^S$}{Spot market electricity quantity [MWh], RES generator}
	\nomenclature[V$Z$]{$\theta_{k\omega},\lambda_k,\mu_{k\omega}, \nu_k$ }{dual variables where $k\in I\cup J$} 
	\printnomenclature
	
\begin{figure}
 \centering
\includegraphics[scale=.580]{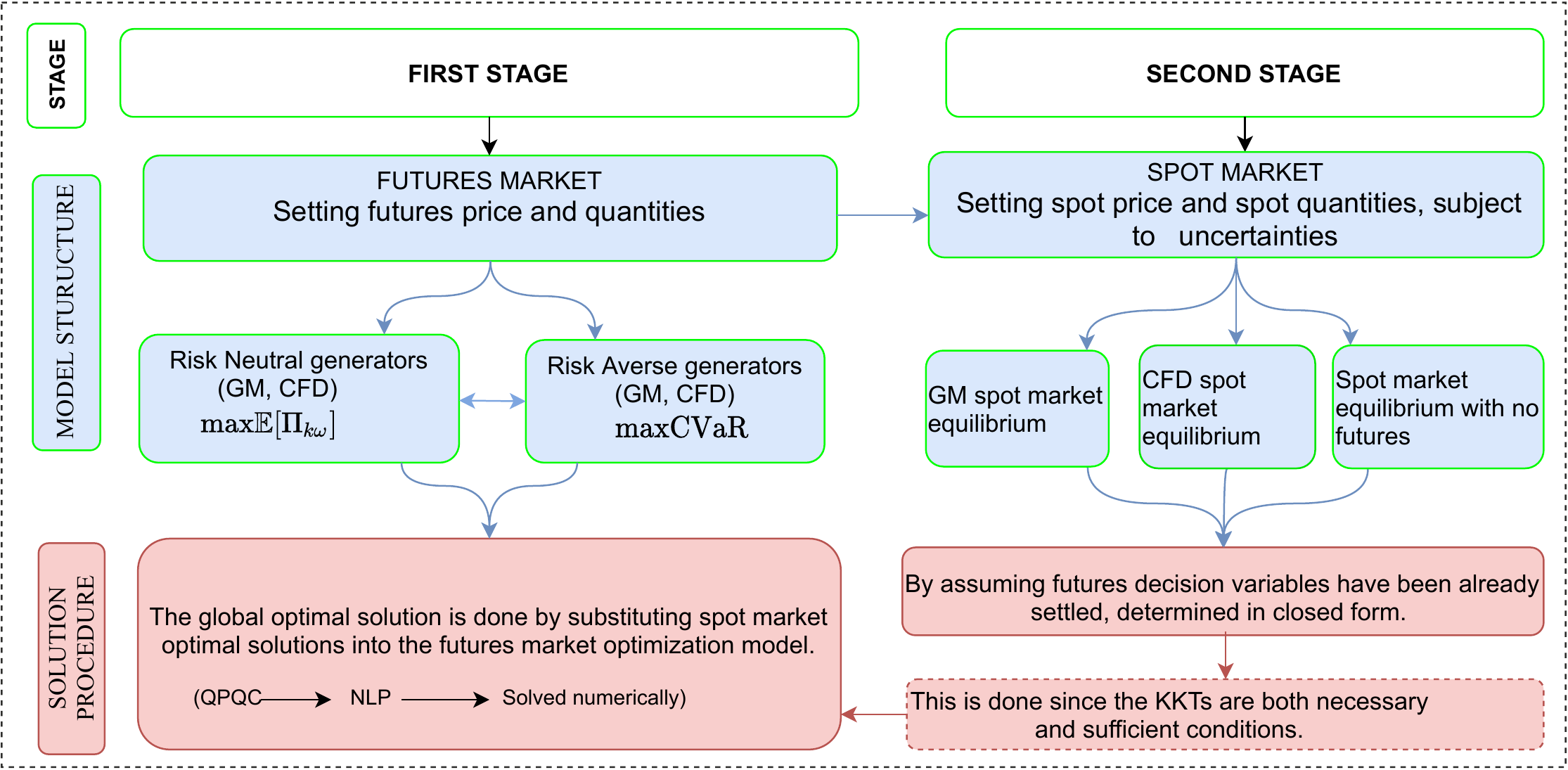}
 \caption{Modeling framework and solution approaches.}%
 \label{Flow_1}%
\end{figure}
\cref{Flow_1} shows the overall modeling framework for the paper. In the spot market, price and quantities are subject to uncertainty. The model formulation starts from the second stage by assuming the first stage variables are already settled. More specifically, the first stage decisions (optimal involvement in the futures market) are taken by anticipating uncertain realizations of RES and spot prices in the second stage and not a given RES generation level. Thus, we explicitly model the uncertainties associated with spot prices and RES quantities, whereby the energy generators use forward trading to hedge the associated profit risk.  In the second stage, the three models (the GM, the CFD, and the spot market with no futures) are obtained in closed form. Then, risk neutral and risk averse generators' problems are solved by substituting the spot market optimal solutions into the futures optimization problem using the CVaR. The problem is a quadratic problem with quadratic constraints (QPQC) so that we transform it into a NLP as proposed in \citet{leyffer2010solving}, which gives an equivalent solution to the original problem. 
 
\subsection{The general model (GM) formulation}
We consider $|I|$ conventional and $|J|$ RES generators producing and trading their electricity in the futures and spot markets. We analyze a two-stage single futures market where both generators can trade their futures quantities ($q_i^F/q_j^F$) at a futures price $P^F$, and they trade the remaining quantities in a subsequent single spot market at a spot price $P_{\omega}^S$, which depends on scenario $\omega$. Conventional generators have a quadratic generation cost function to generate $q_i^F$ and $q_{i\omega}^S$ in the futures and in the spot markets, respectively. Electricity demand, conventional generators' production costs, and availability of RES are the sources of uncertainty in the model. 
Under this market setting, generators have to make their first stage decision (amount of electricity to sell in the futures market) anticipating those uncertainties. These uncertainties are thus characterized by scenarios $\omega\in\Omega$ that represents possible realizations of the final consumers' demand, generation cost, and availability of RES. Given this market setting, conventional generators' profit is expressed as:
\begin{equation}
\Pi_{i\omega}= P^Fq_i^F+P^S_{\omega}q_{i\omega}^S-\left( a_{i\omega}+ b_{i\omega} (q_i^F+q_{i\omega}^S)+\frac{1}{2}c_{i\omega}(q_i^F+q_{i\omega}^S)^2\right)\quad \forall i,\forall\omega\label{L_1}
\end{equation}
where the first term stands for revenue from futures market selling, the second term revenue from spot market selling, and the last term expresses the quadratic conventional generator $i$'s  cost function where $a_{i\omega}\geq 0$, $b_{i\omega}\geq 0$ and $c_{i\omega}\ge 0$ are cost parameters subject to uncertainty. Unlike demand and RES uncertainties, generation cost uncertainty is considered in a few previous studies \citep{mendelson2007strategic,diaz2010electricity,lucheroni2021internal}. Hence, we consider a realistic quadratic cost function for conventional generators where quantity is the sum of the electricity traded in the futures market and the spot market, trading off the computational burden it adds to our model under uncertainty. Conventional generators’ cost uncertainty is mainly caused by fuel (gas, oil, and coal) prices (the interested reader can refer \citet{cain2004implications} and \citet{gautam2007optimal} for quadratic cost structure and cost characteristics).  Futures market decision variables $q_i^F$ and $P^F$ are scenario-independent as the futures market is settled before the uncertain parameters are observed. 

 Similarly, the profit for RES generator with zero generation cost assumption\footnote{This is because once RES generators are built and operational, their marginal generating cost can be assumed to zero. This justifies economically to utilize the entire RES capacity available in the system to satisfy the demand, since that capacity has already been paid for \cite{morales2017inefficiency,pineda2018renewable}. } is expressed as:
\begin{equation}\label{L_2}
\Pi_{j\omega}=P^Fq_j^F+P^S_{\omega}q_{j\omega}^S\quad \forall j,\forall\omega
\end{equation}
where the first term stands for revenue in the futures market and the second term represents the revenue in the spot market. Let's consider $Q_{j\omega}$ as a stochastic parameter representing the total amount of RES generator $j$ production with scenario $\omega$ which is traded in the futures and spot markets. RES generators are nondispatchable so that $Q_{j\omega}=q_j^F+q_{j\omega}^S$, and hence trade $q_{j\omega}^S=Q_{j\omega}-q_j^F$ in the spot market. 
 
 Thus, (\ref{L_2}) can be simplified as $$\Pi_{j\omega}=(P^F-P_{\omega}^S)q_j^F+P^S_{\omega}Q_{j\omega}.$$
The inverse demand curve in the spot market is expressed as:
\begin{equation}
P_{\omega}^S=\gamma_{\omega}^S-\beta_{\omega}^S\left( \sum_{i\in {I}} (q_{i\omega}^S+q_i^F)+\sum_{j\in {J}}(q_{j\omega}^S+q_j^F)\right) \label{L_3} 
\end{equation}
 
with $\gamma_{\omega}^S>0$ and $\beta_{\omega}^S>0$ for the price-demand function to be well-behaved. Since $Q_{j\omega}$ is a parameter, by rearranging, we can simplify (\ref{L_3}) as:
\begin{equation}\label{L_4}
P_{\omega}^{S}=\hat{\gamma}_{\omega}^{S}-\beta_{\omega}^S\sum_{i\in {I}} (q_{i\omega}^S+q_i^F)\quad \forall \omega
\end{equation}
where $\hat{\gamma}_{\omega}^{S}=\gamma_{\omega}^{S}-\beta_{\omega}^S\sum_{j\in {J}} Q_{j\omega}$,   so that $\hat{\gamma}_{\omega}^{S}$ and $\beta_{\omega}^S$ are the component of demand intercept and demand slope, at scenario $\omega$, respectively. Demand uncertainty is represented by $\hat{\gamma}_{\omega}^{S}$ and $\beta_{\omega}^S$, which are scenario-dependent parameters. In (\ref{L_4}), $P_{\omega}^S$ is the spot market price, defined endogenously wherein $\sum_{i\in {I}} (q_{i\omega}^S+q_i^F)$ are the quantities that can be sold in the spot market. Since generators have the option to trade both in the futures and spot markets, they can trade their unsold quantities from the futures market in the spot market. Therefore, both futures and spot quantities affect the spot market clearing price.  
Though it would be realistic to consider the futures market over time (monthly, quarterly) and the spot market over time (day-ahead, real-time), due to the complexity of our model and analytical tractability, we assume a single futures market followed by a subsequent single spot market where the electricity generation and delivery takes place. Besides, instead of taking the usual assumption that the futures price is the expected spot price, we endogenously define it with an inverse demand curve as:
\begin{equation}\label{L_Price}
P^F=\gamma^F-\beta^F\left(\sum_{i\in {I}} q_{i}^F+\sum_{j\in {J}} q_{j}^F\right). 
\end{equation}
 In (\ref{L_Price}), $P^F$ is the price in the futures market, and $\sum_{i\in {I}} q_{i}^F +\sum_{j\in {J}} q_{j}^F$ are futures quantities for conventional and RES generators.Defining the futures price by relaxing the non-arbitrage condition is well-established in the literature as: (1) it is possible to obtain the optimal equilibrium futures price by equating aggregate demand and supply, and without making any assumption to express its relationship with the spot price \citep{oliveira2021analysis}. (2) in non-storable commodities, like electricity, the forward price would be a biased forecast of the spot price due to the presence of electricity forward risk premia \citep{considine2001risk, diko2006risk}. (3) since we are dealing with a single-period spot market, it is not insightful to define futures price as expected spot prices (in our case, we could only have expectation over a finite set of scenarios).  

To characterize the futures market decision variables (stage-one variables), we compute the equilibrium in the spot market (stage-two), assuming the futures market has already settled. As a result, the resulting spot market prices and quantities, per scenario $\omega$, are parametrized with the futures market decisions. To deal with the futures market equilibrium, we move one step backward in time and use the spot market equilibrium decision variables to characterize the global solution.  
\subsubsection{Spot market equilibrium (stage-two)}\label{section_1}
In this section, we analytically derive the equilibrium conditions in the spot market starting from the equilibrium price-demand curve of generators in \cref{L_6} followed by the spot market equilibrium quantities in \cref{L_9}. 
\begin{ass}\label{L_as_2}
	Each generator $k$ has an estimation of the impact that its generation $q_{k\omega}^S, q_k^F$ may have in the spot market price and its rival ($-k$) quantities
	$ P^S_{\omega}=P^S_{\omega}(q_{k\omega}^S, q_k^F)$ and $q_{-k\omega}^S=q_{-k\omega}^S(q_{k\omega}^S)\quad\forall k,\forall\omega.$
\end{ass} 
This is an important assumption to derive both the equilibrium price and quantity in the spot market. The assumption is based on the idea of conjectural variations, an approach that is used widely to study oligopolistic markets. It is used to model an oligopolistic market wherein firms react to price by conjecturing (accurately or not) how changes in their decisions may impact the production decisions of their competitors, which is a generalization of standard Nash–Cournot models to represent imperfect competition \citep{day2002oligopolistic,aid2011hedging,mousavian2020equilibria}.
\begin{prop}\label{L_6}
	Given \cref{L_as_2} and the futures market quantities $q_k^F, \forall k\in I\cup J$, at each scenario $\omega$, the equilibrium spot market price is expressed as:
	\begin{equation}\label{SME_1}
	   P_{\omega}^S=\varphi_{\omega}\left[\hat{\gamma}_{\omega}^{S}-\beta_{\omega}^S\sum_{i\in {I}}q_{i}^F+\beta_{\omega}^S\sum_{i\in {I}}\tau_{i\omega}(b_{i\omega}+c_{i\omega}q_i^F)\right] \quad \forall \omega	
	   \end{equation}
	where $\tau_{i\omega}=\frac{1}{\beta_{\omega}^S(1+\delta_i)+c_{i\omega}}$ and $\varphi_{\omega}=\frac{1}{1+\beta_{\omega}^S\sum_{i\in {I}}\tau_{i\omega}}.$ 
\end{prop}
 \cref{L_6} shows the closed-form solution of the equilibrium spot price (the spot market clearing price) which is expressed in terms of futures market quantities.
The parameter $\delta_i=\sum_{i\ne k}\frac{\partial q_{k\omega}^S}{\partial q_{i\omega}^S}$ measures  the level of competition in the spot market, as analyzed in \citet{lindh1992inconsistency} where $\delta_i=-1$ means generator $i$ behaves as a competitive generator, $\delta_i=0$ means generator $i$ behaves as a Cournot generator, and $\delta_i=m-1$ means that generator $i$ behaves as à la Cournot while being part of a cartel, where $m$ is the number of identical cartel members. The proof for this proposition is found in \ref{Prop_1}.
 
\begin{prop}\label{L_9}
At each scenario $\omega$, the optimal spot quantity $q_{i\omega}^S$ of conventional generator $i$, given the futures market quantity $q_k^F,$ with knowledge of \eqref{L_3}, and $P_{\omega}^S$  in \ref{Prop_1} is expressed as:
	\begin{equation}\label{L_q_1}
q_{i\omega}^S=\tau_{i\omega}\varphi_{\omega}\left[ \hat{\gamma}_{\omega}^{S}-\beta_{\omega}^S\sum_{i\in {I}}q_{i}^F+\beta_{\omega}^S\sum_{i\in {I}}\tau_{i\omega}(b_{i\omega}+c_{i\omega}q_i^F)\right]
	-\tau_{i\omega}\left(b_{i\omega}+c_{i\omega}q_i^F\right). 	
	\end{equation}
\begin{eqnarray}\label{L_c_2}
 q_{j\omega}^S=Q_{j\omega}-q_j^F\quad \forall j,\forall\omega. 
\end{eqnarray}
\end{prop}
The explicit formula for conventional generator $i$'s quantity in the spot market is
	$q_{i\omega}^S=\tau_{i\omega}(P_{\omega}^S-b_{i\omega}-c_{i\omega}q_i^F)$ from \eqref{L_8}. Then, substituting the equilibrium spot price, which is expressed in (\ref{L_edi_3}) into \eqref{L_8} gives the optimal   spot market quantity for conventional generator $i$ ($q_{i\omega}^S$).
Similarly, given $Q_{j\omega}$ i.e., the total RES production, the optimal spot market quantity can be directly derived as (\ref{L_c_2}). 
Since renewables are the first technologies to be dispatched in the spot market based on merit order \cite{morales2017inefficiency,pineda2018renewable,graf2020simplified}, they can only sell at the spot market the energy left after supplying the already signed futures contract  (\ref{L_c_2}). However, their strategic decision in the first stage is taken simultaneously with the conventional generators as expressed in the CVaR formulation in (\ref{L_100}). 

\subsubsection{Futures market analysis (stage-one)}\label{sec_1}
Once the spot market equilibrium conditions are obtained in closed-form, we can go one step backward in time to deal with the futures market equilibrium outcomes using the optimal spot market decision variables ($P_{\omega}^S,q_{i\omega}^S$ and $q_{j\omega}^S$). Then, we characterize the joint maximization of all generators' expected, or risk averse profit (with both technologies).
The backward approach is a standard solution procedure to characterized  two-stage equilibrium models. In this regard, the initial second stage equilibrium can be viewed as the optimal response of the market as a whole, as a function of the first stage variables. Then, taking this response into account, generators decide their optimal first-stage strategy \citep{allaz1993cournot,mousavian2020equilibria}.

To that end, first we start from the conventional generator's profit function expressed in (\ref{L_1}), RES generator's profit expressed in (\ref{L_2}), and compute the derivative with respect to futures market quantities. 
\begin{subequations}\label{L_13}
	\begin{align}
	\frac{\partial \Pi_{i\omega}}{\partial q_i^F}&=\frac{\partial P^F}{\partial q_i^F}q_i^F+P^F+\frac{\partial P_{\omega}^S}{\partial q_i^F}q_{i\omega}^S+P_{\omega}^S\frac{\partial q_{i\omega}^S}{\partial q_i^F}
	-b_{i\omega}\left(1+\frac{\partial q_{i\omega}^S}{\partial q_i^F} \right)+\nonumber\\
	&-c_{i\omega}\left[ q_i^F+q_{i\omega}^S \right] \left(1+\frac{\partial q_{i\omega}^S}{\partial q_i^F} \right)\forall i,\forall\omega\\ 
	\frac{\partial \Pi_{j\omega}}{\partial q_j^F}&=\left(\frac{\partial P^F}{\partial q_j^F}-\frac{\partial P_{\omega}^S}{\partial q_j^F} \right)q_j^F+P^F-P_{\omega}^S+\frac{\partial P_{\omega}^S}{\partial q_j^F}Q_{j\omega}\quad \forall j, \forall\omega. 
	\end{align}
\end{subequations}
\begin{ass}\label{L_as_l}
	$\forall k \in I\cup J$, each generator $k$ has an estimation of the impact its generation $q_{k}^F$ may have in the futures price and competitors' generation, where $ P^F=P^F(q_{k}^F)$ and $q_{-k}^F=q_{-k}^F(q_{k}^F)$. 
\end{ass}
This is considered to compute $\frac{\partial \Pi_{k\omega}}{\partial q_{k}^F}$, which now depends on $\psi_k=\frac{\partial q_{-k}^F}{\partial q_k^F}$. Therefore, the different levels of competition in the futures market can be modeled as: $\psi_k=-\frac{1}{I+J-1}$ for competitive generator, $\psi_k=0$ for Cournot generator, and $\psi_k>0$ for generator part of a cartel.
From (\ref{L_13}), we can observe there are several partial derivatives ($\frac{\partial P^F}{\partial q_i^F},\frac{\partial P_{\omega}^S}{\partial q_i^F}$ and $\frac{\partial q_{i\omega}^S}{\partial q_i^F}$ (for conventional generators), and $\frac{\partial P^F}{\partial q_j^F},\frac{\partial P_{\omega}^S}{\partial q_j^F}$ and $\frac{\partial q_{j\omega}^S}{\partial q_j^F}$ (for RES generators), which are derived using (\ref{L_Price}), (\ref{L_edi_3}), (\ref{L_q_1}) and (\ref{L_c_2})) that can influence the futures market (the global solution in the CVaR formulation) as we will see in the next section.

Given the equilibrium spot market outcomes, the partial derivatives needed to be computed for both sets of generators are:
\begin{subequations}\label{L_15}
	\begin{align}
	\frac{\partial P^F}{\partial q_i^F}&=-\beta^F\left( 1+\sum_{i\ne k}\frac{\partial q_{k}^F}{\partial q_i^F}+\sum_{j\in {J}}\frac{\partial q_{j}^F}{\partial q_i^F}\right)=-\beta^F(1+(I+J-1)\psi_i)\quad \forall i \label{Par_1} \\
	\frac{\partial P_{\omega}^S}{\partial q_i^F} &=\varphi_{\omega}\left[ \left( -\beta_{\omega}^S(1+(I-1)\psi_i)\right)\right]  + 
	\varphi_{\omega}\left[ \beta_{\omega}^Sc_{i\omega}\tau_{i\omega}+\beta_{\omega}^S\sum_{k\ne i}^{I}c_{k\omega}\psi_k\tau_{k\omega}\right]\quad \forall i, \forall\omega \\
	\frac{\partial q_{i\omega}^S}{\partial q_i^F} &=\tau_{i\omega}\left(\frac{\partial P_{\omega}^S}{\partial q_{i}^F}-c_{i\omega} \right)\quad \forall i, \forall\omega\\
	\frac{\partial P^F}{\partial q_j^F} &=-\beta^F\left( 1+(I+J-1)\psi_j\right) \quad \forall j \\
	\frac{\partial P_{\omega}^S}{\partial q_j^F} &=0\quad \forall j,\forall\omega \\
	\frac{\partial q_{j\omega}^S}{\partial q_j^F} &=-1\quad \forall j,\forall\omega\label{Par_end}. 
	\end{align}
\end{subequations}
\subsubsection{Futures market equilibrium under the CVaR}\label{sec_2}
By replacing the equilibrium spot price $P^S_{\omega}$ and quantities ${q^S_{i\omega}}$ and ${q^S_{j\omega}}$ in the profit functions $\Pi_{i\omega}$ and $\Pi_{j\omega}$, we can obtain parameterized profits in the futures market decision variables:
\begin{subequations}
	\begin{align}
	&\Pi_{i\omega}=\Pi_{i\omega}(q^F_i,q^F_{-i},q^F_{j\in J})\quad \forall i, \forall\omega\\
	&\Pi_{j\omega}=\Pi_{j\omega}(q^F_j,q^F_{-j},q^F_{i\in I})\quad \forall j, \forall\omega.
	\end{align}
\end{subequations}
The risk aversion level of generators is considered by using the CVaR measure. CVaR is a coherent\footnote{A functional $\mathcal{R}:\mathcal{L}^2\rightarrow(\infty,\infty]$ is called a coherent measure of risk in the extended sense if (1) $\mathcal{R}(C)=C$ for all constants $C$,
	(2) $\mathcal{R}((1-\lambda)X +\lambda X') \le(1-\lambda)\mathcal{R}(X)+\lambda \mathcal{R}(X)$ for $\lambda \in(0,1)$ ("convexity"),\\
	(3) $\mathcal{R}(X)\le \mathcal{R}(X')$ ("monotonicity"),
	(4) $\mathcal{R}(X)\le 0$ when $||X^k-X||_2 \rightarrow 0$ ("closedness"),
	(5) $\mathcal{R}(\lambda X)=\lambda\mathcal{R}(X)\text{for}\lambda>0$ ("positive homogeneity").} 
risk measure that calculates VaR and optimizes CVaR simultaneously. 
The CVaR at $\alpha$ confidence level can be defined as the expected value of the profit smaller than the ($1-\alpha$)-quantile of the profit distribution. In the scenario-based stochastic optimization method, CVaR$_{\alpha}$ measures the expected profit in the $(1-\alpha)\times 100\%$ worst scenarios. 
The $(1-\alpha)-$quantile of the profit distribution is known as VaR, which is the largest value ensuring that the probability of obtaining a profit less than that value is lower than $1-\alpha$, $\forall\alpha\in[0,1]$. 
Thus, the profit maximization problem solved by the risk (neutral and averse) generators is formulated in \eqref{L_100}, where $\sigma_{k\omega}$ is the probability assigned by generator $k$ to scenario $\omega$, and $1-\alpha$ represents the level of significance associated with the CVaR. 
The risk averse problem solved by each generator is \citep{rockafellar2000optimization}:
\begin{subequations}\label{L_100}
	\begin{align}
	\displaystyle \max_{\xi_k, \eta_{k\omega}, q_k^F} & {(1-\phi)\sum_{\omega\in {\Omega}}\sigma_{k\omega}\Pi_{k\omega}}+\phi \left[\xi_k-\frac{1}{ 1-\alpha}\sum_{\omega\in {\Omega}}\sigma_{k\omega}\eta_{k\omega} \right]\\
	&\textrm{s.t.}\nonumber\\
	&\eta_{k\omega}+\Pi_{k\omega}-\xi_k\geq 0 \quad {:\mu_{k\omega}} \quad\forall\omega\label{L_Cv_2}\\
	& \eta_{k\omega}\geq 0\quad {:\theta_{k\omega}} \quad \forall\omega\label{L_Cv_3}\\
	& q_{k}^{F_{min}}\leq q_{k}^{F}\leq q_{k}^{F_{max}}\quad {:\nu_{k}^{min},\nu_{k}^{max}} \quad\forall k\label{L_d_2}
	\end{align}
\end{subequations}
The objective function is represented by the expected profit (first term for risk neutral generators) and the CVaR (second term for risk averse generators) multiplied by a factor $\phi\in [0,1]$ that regulates a trade-off between the expected profit and the CVaR for a given level $\alpha$. In this setting, $\phi=0$ corresponds to a risk neutral generator (maximization of the expected profits), and $\phi=1$ to the most risk averse setting in which all the weight in the CVaR is considered,  as depicted in the modeling framework in \cref{Flow_1}.  Parameter $\sigma_{k\omega}$ represents the scenario occurrence probability which is assumed to be $\frac{1}{\Omega}$. At the optimal solution, $\xi_{k}$ represents the VaR, and the auxiliary variable $\eta_{k\omega}$ equals (for each scenario) the positive difference between the VaR and the profit $(\Pi_{k\omega})$, which is expressed with constraint (\ref{L_Cv_2}).Constraints (\ref{L_d_2}) enforce the lower and upper bounds for electricity schedule of generators  (both conventional and RES generators as $k=I\cup J$). Since both generators have capacity constraints (between zero and $q_{k}^{F_{max}}$), they decide their production based on their benefits.  The dual variables associated with constraints (\ref{L_Cv_2}), (\ref{L_Cv_3}) and (\ref{L_d_2}) are $\mu_{k\omega},\theta_{k\omega},\nu_{k}^{min}$, and $\nu_{k}^{max}$, which are treated as variables in the model.
 
The equilibrium is then obtained by solving simultaneously (\ref{L_100}) for all generators. This is done by replacing \eqref{L_100} by its associated Karush-Kuhn-Tucker (KKT) system of optimality conditions. Therefore, the KKT system associated with each generator is:
\begin{subequations}\label{L_eq:5}
	\begin{align}
	&\frac{\partial\mathcal{L}}{\partial q_k^F}= {-(1-\phi)\sum_{\omega\in {\Omega}}\sigma_{k\omega}\frac{\partial\Pi_{k\omega}}{\partial q_k^F}}-\sum_{\omega\in {\Omega}}\mu_{k\omega}\frac{\partial\Pi_{k\omega}}{\partial q_k^F}-\nu_k^{min}+\nu_k^{max}=0 \quad\forall k\label{L_second_a}\\
	&\frac{\partial\mathcal{L}}{\partial \eta_{k\omega}}=\phi\frac{1}{1-\alpha}\sigma_{k\omega}-\mu_{k\omega}-\theta_{k\omega}=0 \quad \forall k,\forall \omega\label{L_second_b}\\
	& \frac{\partial\mathcal{L}}{\partial \xi_{k}}=-\phi + \sum_{\omega}^{\Omega}\mu_{k\omega}=0\quad \forall k\label{L_second_c}\\
	&(0\leq \eta_{k\omega}+\Pi_{k\omega}-\xi_k) \perp (\mu_{k\omega}\geq 0) \quad \forall k,\forall \omega\label{L_second_d}\\
	&(0\leq \eta_{k\omega}) \perp (\theta_{k\omega}\geq 0)\quad \forall k,\forall \omega\label{L_second_e}\\
	&(0\leq q_{k}^{F}-q_{k}^{F_{min}}) \perp (\nu_{k}^{min}\geq 0)\quad \forall k\label{L_second_f}\\
	&(0\leq q_{k}^{F_{max}}-q_{k}^{F}) \perp (\nu_{k}^{max}\geq 0)\quad \forall k.\label{L_second_g}
	\end{align}
\end{subequations}
The complementarity conditions are denoted by $(0\le x)\perp (y\ge0)$ which is equivalent to: $x\ge0,y\ge0$ and $xy=0.$    One way to solve system \eqref{L_eq:5} for all $k$ is by using the technique discussed in \citet{leyffer2010solving} where a nonlinear programming reformulation of KKT systems is proposed. This technique is based on minimizing the sum of the complementarity products subject to the remaining KKT conditions. Solutions where the objective function is zero guarantee that all the original KKT conditions are met,   as zero is the lower bound for the objective function (sum of nonnegative terms).   Thus, we can consider the problem in the following extended form:

\begin{subequations}\label{L_48}
	\begin{align}
	\displaystyle \min & \sum_{i\in {I}}\sum_{\omega\in {\Omega}}\left[ \mu_{i\omega}(\eta_{i\omega}+\Pi_{i\omega}-\xi_{i})+\eta_{i\omega}\theta_{i\omega}\right]	
	 +\sum_{j\in {J}}\sum_{\omega\in {\Omega}}\left[ \mu_{j\omega}(\eta_{j\omega}+
 \Pi_{j\omega}-\xi_{j})+ \eta_{j\omega}\theta_{j\omega}\right]+\nonumber\\
	&+\sum_{i\in {I}}\left[\left(q_i^F-q_{i}^{F_{min}}\right)\nu_{i}^{min}+ \left(q_{i}^{F_{max}}-q_i^F \right)\nu_{i}^{max}\right]\nonumber\\
	&+\sum_{j\in {J}}\left[\left(q_j^F-q_{j}^{F_{min}} \right)\nu_{j}^{min}+\left(q_{j}^{F_{max}}-q_j^F\right)\nu_{j}^{max}\right]\\
	&\textrm{subject to}\nonumber\\
	&\text{equalities}\quad\eqref{L_second_a}-\eqref{L_second_c}\\
	&\text{partial derivatives}\quad\eqref{Par_1}-\eqref{Par_end}\\
	&\text{inequalities}\quad\eqref{L_second_d}-\eqref{L_second_g}\\
	& (\ref{L_1}),(\ref{L_2}),(\ref{SME_1}), (\ref{L_q_1}), (\ref{L_c_2})
	\end{align}
\end{subequations}
where the objective function is formulated by the sum of the complementarity products all the remaining KKT optimality conditions entered as constraints. (\ref{L_1}) and (\ref{L_2}) are the profit definitions of the conventional and RES generators, respectively.    (\ref{Par_1})-(\ref{Par_end}) are the partial derivatives with respect to the futures market quantities. Finally, (\ref{SME_1}) is equilibrium price in the spot market, whereas (\ref{L_q_1}) and (\ref{L_c_2}) are the equilibrium quantities of conventional and RES generators, respectively. Note that the global NLP presented in (\ref{L_48}) represents the joint solution of several quadratic problems with quadratic constraints (QPQC), one per player.
 
\subsection{Contracts for differences (CFD) model formulation} 
Our second model deals with a futures market which settles financially based on contracts for differences (CFD), that can be used by generators to protect themselves from the price and quantity fluctuations that arises in the spot electricity market \citep{kristiansen2004pricing,oliveira2013contract}.
The structure of the electricity market with CFD is similar to the market in the GM except in the CFD generators use the strike price as a reference price and there is a financial settlement at expiry rather than a physical delivery.
For all generators $k$, the contract signed is a CFD that fixes the electricity amount $q_k^F$, and the delivery price $P^F$ (the strike price) in the first stage. When the spot market takes place, the spot price $P_{\omega}^S$ can be either greater than or less than the futures price $P^F$. Therefore, if the spot market price $P_{\omega}^S$ is lower than $P^F,$ generator $k$ earns the difference between the two price times the amount of electricity agreed in the contract $(P^F-P_{\omega}^S)q_k^F$. Otherwise, generator $k$ will pay the difference between the two prices times the amount of energy agreed in the contract at time-one $(P_{\omega}^S- P^F) q_k^F$, if the spot market price surpasses the futures market price at the spot market.
Therefore, profit for conventional generator is expressed as:
\begin{equation}\label{L_cfd_1}
\Pi_{i\omega}=(P^F-P_{\omega}^S)q_i^F+P_{\omega}^Sq_{i\omega}^S-a_{i\omega}-b_{i\omega}q_{i\omega}^S-\frac{1}{2}c_{i\omega}(q_{i\omega}^S)^2\quad \forall i, \forall \omega
\end{equation} and 
\begin{equation}\label{L_CFD_1}
\Pi_{j\omega}=(P^F-P_{\omega}^S)q_j^F+P_{\omega}^Sq_{j\omega}^S\quad \forall j, \forall \omega
\end{equation} for RES generator. The profit formulations in the CFD are different from the GM profit formulations for conventional generators in two respects. First, unlike the GM, conventional generators' revenue in the futures market with CFD is expressed as price difference times quantity, $(P^F-P_{\omega}^S)q_i^F$ and second, the production cost function depends only on the spot quantity ($q_{i\omega}^S$), as there is no electricity generation in the futures market in the CFD. Similarly, RES generators' profit is different, though the first term seems similar to the GM due to a rearrangement for convenience. So that the second term in (\ref{L_CFD_1}) depends only on the spot quantity, rather than $P_{\omega}^SQ_{j\omega}^S$ as in the GM. 
The analytical derivation follows a similar procedure then in \Cref{section_1}, where we need to start from the spot market and then come back to the futures market to compute the market equilibrium.
\subsubsection{Spot market equilibrium (stage-two)}
Since we do not have futures production ($q_i^F$ and $q_j^F$ are financial quantities) in the CFD, the spot market price is described as:
\begin{equation}\label{L_222}
P_{\omega}^{S}=\gamma_{\omega}^{S}-\beta_{\omega}^S\sum_{i\in {I}} q_{i\omega}^S-\beta_{\omega}^S\sum_{i\in {I}} Q_{j\omega}=\hat{\gamma}_{\omega}^{S}-\beta_{\omega}^S\sum_{i\in {I}} q_{i\omega}^S\quad \forall\omega.
\end{equation}
\begin{prop}\label{cdf_1}
Taking definition of $\varphi_{\omega}$ and $\tau_{i\omega}$ from \cref{L_6}, the equilibrium spot price and spot quantities in the CFD are described as:
\begin{equation}
	P_{\omega}^{S}=\varphi_{\omega}\left( \hat{\gamma}_{\omega}^{S}+\beta_{\omega}^S\sum_{i\in {I}}\tau_{i\omega}b_{i\omega}-\beta_{\omega}^S\sum_{i\in {I}}q_i^F \beta_{\omega}^S(1+\delta_i)\tau_{i\omega}\right)\quad\forall \omega 
	\end{equation}
	and
	\begin{equation}
	q_{i\omega}^S=\tau_{i\omega} \left( \varphi_{\omega}[ \hat{\gamma}_{\omega}^{S}+\beta_{\omega}^S\sum_{i\in {I}}\tau_{i\omega}b_{i\omega}-\beta_{\omega}^S\sum_{i\in {I}}q_i^F \beta_{\omega}^S(1+\delta_i)\tau_{i\omega}]
	-b_{i\omega} +  q_i^F \beta_{\omega}^S(1+\delta_i)\right) \quad\forall i, \forall\omega.
		\end{equation}
\end{prop}
 
The proof for \cref{cdf_1} is found in \ref{Prop_2}. As we can see, the optimal spot market clearing price and spot market quantities for generator $i$ are analytically derived in closed form, which are again expressed in terms of futures market decision variables.
 
\subsubsection{Futures market analysis (stage-one)}
Similar to the GM, the partial derivatives with respect to futures quantities, needed to characterize conventional generators and RES generators in CFD, are expressed as follows:
\begin{equation}\label{L_620}
\frac{\partial\Pi_{i\omega}}{\partial q_i^F}=\frac{\partial P^F}{\partial q_i^F}q_i^F+P^F-\frac{\partial P_{\omega}^S}{\partial q_i^F}q_i^F-P_{\omega}^S+\frac{\partial P_{\omega}^S}{\partial q_i^F}q_{i\omega}^S\nonumber
+\frac{\partial q_{i\omega}^S}{\partial q_i^F}P_{\omega}^S-b_{i\omega}-c_{i\omega}q_{i\omega}^S
\end{equation}
\begin{equation}\label{L_621}
=\left( \frac{\partial P^F}{\partial q_i^F}-\frac{\partial P_{\omega}^S}{\partial q_i^F}\right) q_i^F+q_{i\omega}^S\left( \frac{\partial P_{\omega}^S}{\partial q_i^F}-c_{i\omega}\right)+
P_{\omega}^S\left(\frac{\partial q_{i\omega}^S}{\partial q_i^F}-1 \right)+P^F-b_{i\omega}\quad \forall i,\forall\omega
\end{equation} 
and
\begin{equation}\label{L_CFD_3}
\frac{\partial \Pi_{j\omega}}{\partial q_j^F}=\left(\frac{\partial P^F}{\partial q_j^F}-\frac{\partial P_{\omega}^S}{\partial q_j^F} \right)q_j^F+P^F-P_{\omega}^S+\frac{\partial P_{\omega}^S}{\partial q_j^F}Q_{j\omega}\quad \forall j,\forall\omega,
\end{equation}
respectively.

By recalling the futures market price expressed in \eqref{L_Price} and considering \cref{L_as_l}, the partial derivatives needed to complete \eqref{L_620} and \eqref{L_CFD_3} are:
\begin{subequations}
	\begin{align}\label{L_74}
	\frac{\partial P^F}{\partial q_i^F}&=-\beta^F(1+(I+J-1)\psi_i) \quad \forall i \\
	\frac{\partial P^F}{\partial q_j^F} &=-\beta^F\left( 1+(I+J-1)\psi_j\right)\quad \forall j  \\
	\frac{\partial P_{\omega}^S}{\partial q_i^F}&=\varphi_{\omega}[  -\beta_{\omega}^S\beta_{\omega}^S(1+\delta_i)\tau_{i\omega} - \beta_{\omega}^S\sum_{i\ne k}^{}\beta_{\omega}^S\psi_{i}(1+\delta_{i})\tau_{i\omega}]\quad \forall i,\forall\omega\\
	\frac{\partial q_{i\omega}^S}{\partial q_i^F}& =\tau_{i\omega}\frac{\partial P_{\omega}^S}{\partial q_i^F} + \tau_{i\omega}\beta_{\omega}^S(1+\delta_i)\quad \forall i,\forall\omega\\
	\frac{\partial P_{\omega}^S}{\partial q_j^F} &=0\quad \forall j,\forall \omega \\
	\frac{\partial q_{j\omega}^S}{\partial q_j^F} &=-1 \quad \forall j,\forall\omega. 
	\end{align}
\end{subequations}
\subsubsection{Risk-averse futures market equilibrium under the CVaR}\label{sec_6}
Once we know the equilibrium spot price $P^S_{\omega}$ and quantities, ${q^S_{i\omega}}$ and ${q^S_{j\omega}}$, we can obtain parameterized the total profits in the futures decision variables:
\begin{subequations}
	\begin{align}
	&\Pi_{i\omega}=\Pi_{i\omega}(q^F_i,q^F_{-i},q^F_{j\in J})\quad \forall i, \forall\omega\\
	&\Pi_{j\omega}=\Pi_{j\omega}(q^F_j,q^F_{-j},q^F_{i\in I})\quad \forall j \forall\omega.
	\end{align}
\end{subequations}
Each player $k \in I \cup J$ maximizes its total profit (risk neutral, or risk averse) by solving an optimization problem similar to \eqref{L_100} in the futures market considering the CVaR. The market equilibrium can be done by following the procedure described in \Cref{sec_2} and concatenate the KKT conditions for each generator to solve with an equivalent NLP formulation.
\subsection{Spot market without the presence of futures contracts}
Finally, we derive the equilibrium market outcomes of a market without the presence of futures trading (a single spot market) for comparison purposes. Thus, the profit for conventional generator in the only spot configuration is expressed as:
\begin{equation}\label{P_d_1}
\Pi_{i\omega}=P^S_{\omega}q_{i\omega}^S-a_{i\omega}- b_{i\omega} q_{i\omega}^S-\frac{1}{2}c_{i\omega}(q_{i\omega}^S)^2 \quad\forall i,\forall\omega
\end{equation}
and the profit for generator $j$ as: 
\begin{equation}\label{P_d_2}
\Pi_{j\omega}=P_{\omega}^SQ_{j\omega} \quad \forall\omega.
\end{equation}
 
\begin{prop}\label{som}
Given \cref{L_as_2} and the profit definitions expressed in (\ref{P_d_1}) and (\ref{P_d_2}), the spot market equilibrium outcomes are analytically derived as:
\begin{equation}
  P_{\omega}^S=\varphi_{\omega}\left[\gamma_{\omega}^{S}-\beta_{\omega}^S\sum\limits_{i=1}^J Q_{j\omega}+\beta_{\omega}^S\sum_{i\in {I}}\tau_{i\omega}b_{i\omega}\right] \quad \forall\omega.  
\end{equation}
\begin{equation}
q_{i\omega}^S=\tau_{i\omega}\varphi_{\omega}\left[\hat{\gamma}_{\omega}^{S}+\beta_{\omega}^S\sum_{i\in {I}}\tau_{i\omega}b_{i\omega}\right]-\tau_{i\omega}b_{i\omega}\quad \forall i,\forall\omega.  
\end{equation}
\end{prop} The proof for \cref{som} is found in the \ref{Prop_3}. As it can be seen, both price and quantity are expressed in terms of parameters, as there are no futures decision variables in this case.
 
\section{Numerical results}\label{L_1203}
In this section, we analyze and discuss numerical examples using calibrated data from the Spanish electricity market that combines futures contracts and pool market structures.  The reasons for us to choose this market are: first, in the Spanish electricity market, generators can sign physical or financial futures contracts. Then, they participate in a daily market, and the market operator maintains the bid-offer balance of electricity power to determine the market price as well as the generation of electricity quantities corresponding to each generator for each hour in the schedule \citep{OMIE2020}.  Second, a large proportion of the electricity is traded by very few large generators whose total output is the sum of several units with different technologies. This is very important from a computational point of view since increasing the number of generators increases the computational burden, as our problem is highly nonlinear. 
 
\subsection{Data}
Therefore, we consider three conventional oligopolistic generators and one RES generator who can produce and retail their generation in the futures and spot markets. \cref{tab_1} presents the expected values of the calibrated data for the cost, demand and RES parameters along with the corresponding standard deviations. From the cost parameters, $a_{i\omega}$ are fixed to arbitrary values $a_{1\omega}=a_{2\omega}=a_{3\omega}=0$, as they do not have a direct influence on the market results. The values of $b_{i\omega}$ and $c_{i\omega}$ are randomly generated by estimating their expected values, which are close to the real market values (see \cite{diaz2010electricity,ruiz2012equilibria,oliveira2013contract,mousavian2020equilibria} for similar approach). Each realization of $b_{i\omega}$ is generated from a multivariate normal distribution with mean $\mu_b=[37, 40, 43]$ [\euro/MWh] and standard deviation $\sigma_b=[3.5, 4.55, 5.59]$ [\euro/MWh], which is calculated using 9\% coefficient of variation (CV), as $\sigma= \mu\times CV$. Similarly, $c_{i\omega}$ is generated randomly from a normal distribution  with $\mu_c=[0.013,0.003,0.019]$ [\euro/MWh$^2$] and standard deviation $\sigma_c=[0.000125,0.0002,0.000195]$ [\euro/MWh$^2$], calculated using 5\% of CV. $150$ and $200$ equiprobable scenarios are considered for the risk neutral case and for the CVaR, respectively. The significance level is fixed at $\alpha=0.90.$ 
\begin{table}
	\centering
	\caption{Mean, CV and standard deviation used to calibrate the data applied in our simulations for conventional generators and the RES parameter. }
	\begin{tabular}{lclll|cll}
		\hline\hline
		Parameters & \multicolumn{1}{l}{$i=1$} & $i=2$ & $i=3$ & $CV$ & \multicolumn{1}{l}{Parameters} & mean & $\sigma$ \\ \hline
		$a_{i\omega}$ & 0.00     & 0.00     & 0.00     & -    & $\gamma^F$  & 180.00         & 18 \\
		$b_{i\omega}$ & 37.00    & 40.00    & 43.00 & 0.09 & $\beta^F$  & 0.005      & 0.0005 \\
		$c_{i\omega}$ & 0.013 & 0.003 & 0.019 & 0.05 & $Q_{j\omega}$ & 0-10,000& 1000 \\ \hline\hline
	\end{tabular}
	\label{tab_1}
\end{table}
On the other hand, the demand curve parameters are generated by approximating the aggregated step-wise demand curve in the spot market using the mean futures market intercept $\mu_{\gamma^F}=180$ and mean slope $\mu_{\beta^F}=0.005.$ We assume that the parameters of the inverse demand curve in the spot market equal, on expectation, those parameters of the futures inverse demand, i.e, $\gamma^F=E[\gamma_{\omega}^S]$ and $\beta^F=E[\beta_{\omega}^S]$. Note that the uncertainty associated with consumer's behavior is represented by both the demand intercept ($\gamma_{\omega}^S$) and its slope ($\beta_{\omega}^S$), which are scenario dependent. Thus, the value of $\gamma_{\omega}^S$ for each scenario is randomly generated with a normal distribution so that $\gamma_{\omega}^S\sim N(\mu_{\gamma^F},\sigma_{\gamma^F})$ and $\beta_{\omega}^S\sim N(\mu_{\beta^F},\sigma_{\beta^F})$, where $\sigma_{\gamma^F}$ and $\sigma_{\beta^F}$ are calculated as $\sigma= \mu\times CV$, assuming CV to be $10\%$ for the corresponding mean values of both parameters. Total RES parameter is randomly generated with its expected value at $5,000$MWh and a standard deviation of 1000 with CV of $20\%$. 

The simulation is done for the RES parameter that ranges from $0$ to $10,000$ MWh (we consider the amount of RES from 0-10,000MWh in a range of 1000MWhs given the expectation of RES to be integrated into the system is increasing realistically). The maximum amount that conventional generators are allowed to generate in the model is arbitrarily set at 6,000, 7,000, and 5,000MWh for the three generators, respectively.    
The RES generator capacity is fixed to an increasing value within the range (0, 10,000) MWh in each iteration.

The model is implemented in JuMP version 0.21.1 \citep{ dunning2017jump} under the open-source Julia language version 1.5.2 \citep{bezanson2017julia}. We use Artelys Knitro solver version 12.2 \citep{knitro} on a CPU E5-1650v2@3.50GHz and 64.00 GB of RAM running workstation. Since there are multiple simulation cases, we report the computational performance of one case as a representative reference: For instance, the overall computational time for the CVaR definition with $|\Omega|=200$ equiprobable scenarios is 917.177904 seconds (156.46 M allocations: 3.199 GiB, 0.10\% gc time).

According to the specification of problem (\ref{L_100}), two risk profiles ($\phi=0$ for the risk neutral and $\phi=1$ for the risk averse generators) are considered. 
We analyze how the relationship between futures and spot markets, risk aversion, and level of competition are affected with respect to high RES penetration (level of renewables). The electricity market outcomes (electricity price, quantity, and generators' profit) are discussed for two market structures (Cournot and perfect competition) and for two models (GM and CFD) with respect to RES penetration. The x-axis in all the plots, but one represents RES penetration that ranges from 0 to 10,000WMh, whereas the y-axis may be price, quantity, or profit.
\subsection{Risk averse generators’ numerical results (\texorpdfstring{$\phi=1$)}{Lg}}
For the risk averse generators, the numerical results are obtained by simulating the model with the parameter $\phi=1$. The results are futures and spot markets equilibrium outcomes when generators maximize their CVaR from our spectral risk measure representation. We analyze the influence of generators' risk aversion behavior on equilibrium market outcomes based on levels of competitiveness (Cournot and perfect competition) under two contracts (physical, or financial).
\begin{figure}
	\centering
	\includegraphics[width=\textwidth]{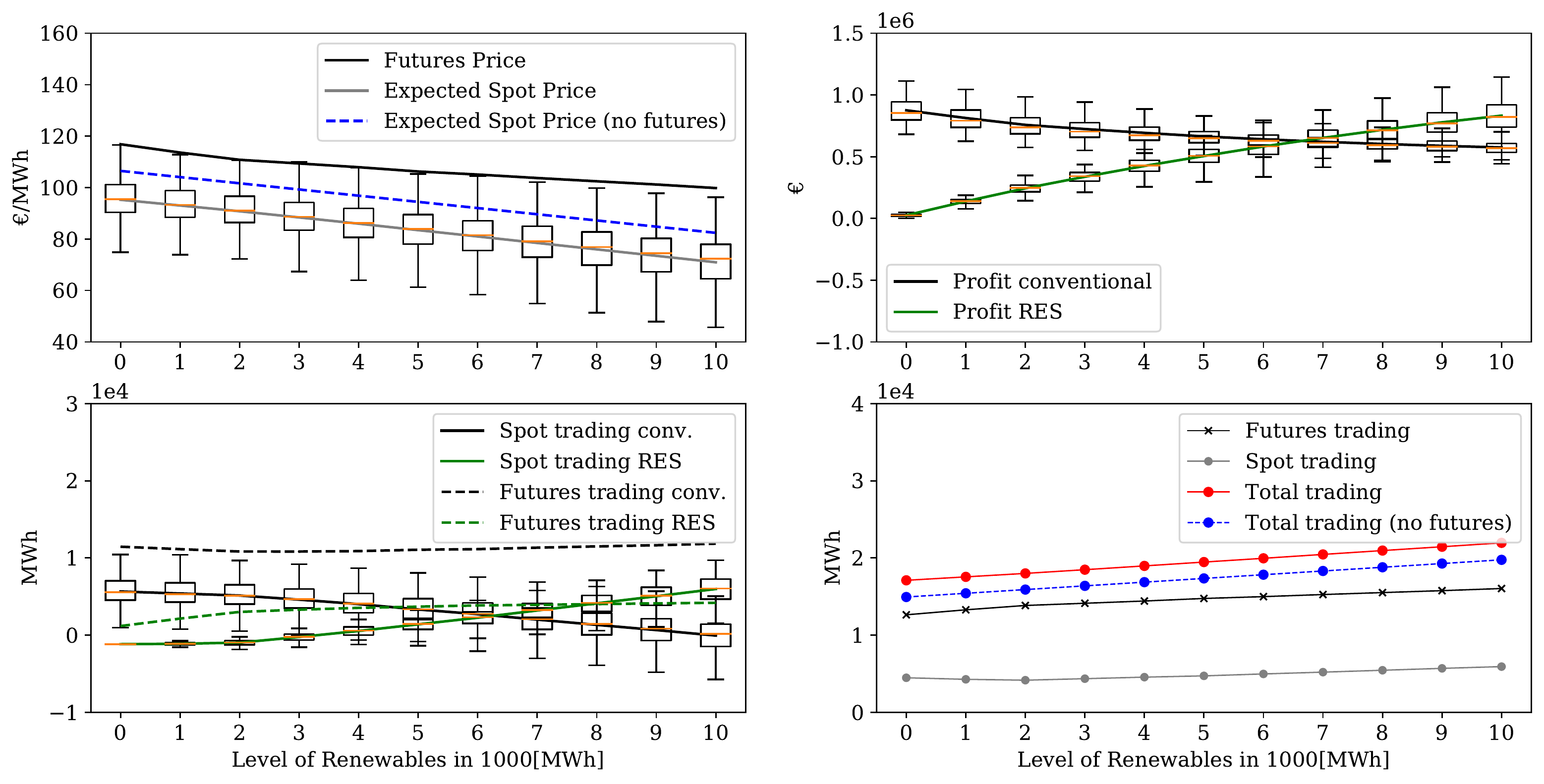}
	\caption{Risk averse Cournot model in the GM}
	\label{L_f1}
\end{figure}
\begin{figure}
	\centering
	\includegraphics[width=\textwidth]{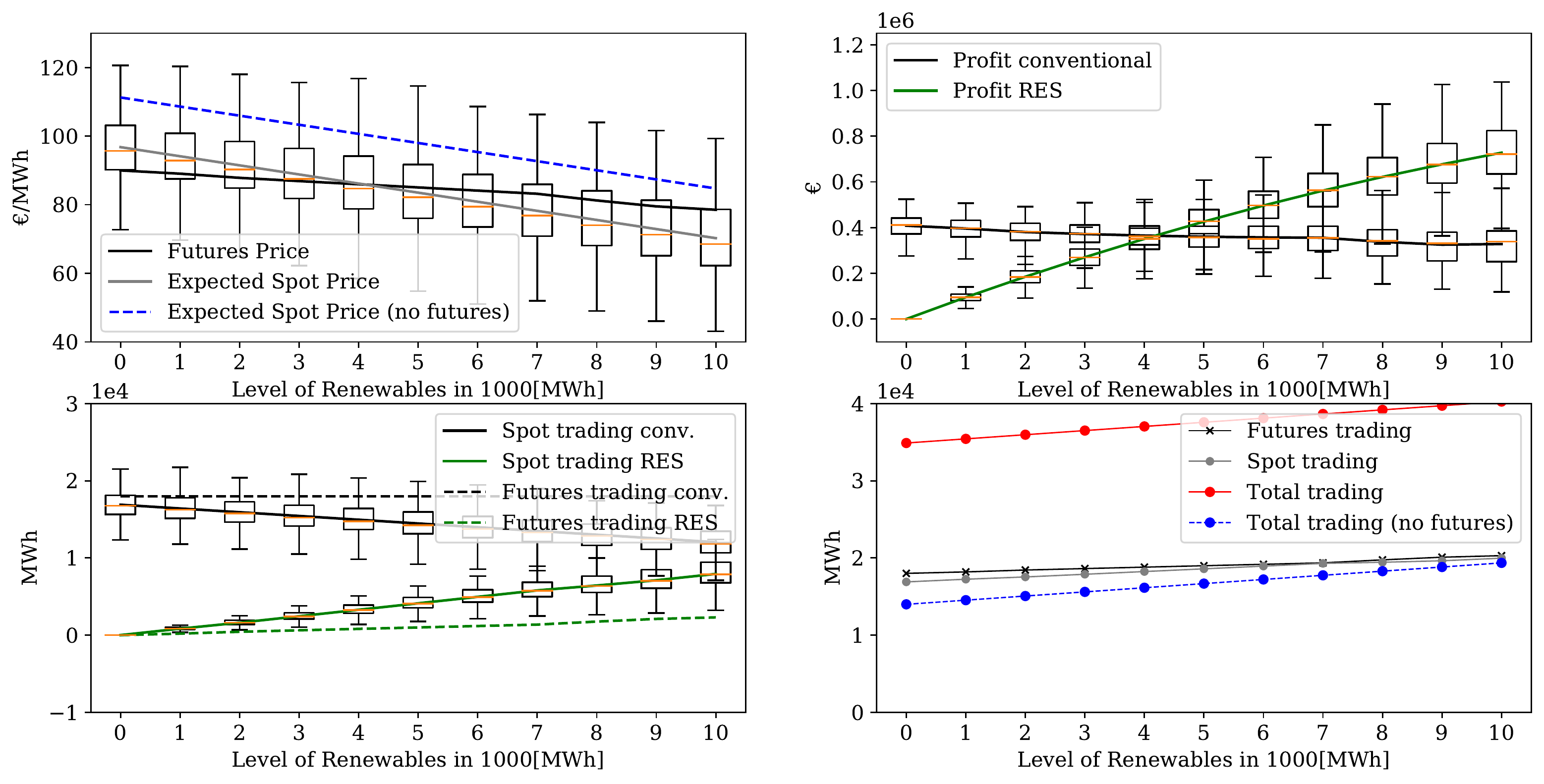}
	\caption{Risk averse Cournot CFD}
	\label{L_f2}
\end{figure}

Let’s examine the results starting from the Cournot model in the general model. Looking at \cref{L_f1}, \cref{L_f2}, and \cref{tab_2}, overall price decreases (downward sloping) with respect to RES penetration. That is because when RES penetration increases, conventional generation with generation cost decreases and is replaced by RES generation (with no generation cost), which may transfer part of the cost saved to consumers. If we look at the general model results, we see that consumers must pay as much as 116\euro/MWh with zero level of renewables, and as low as 101\euro/MWh in the futures market with higher (10,000MWh) RES penetration. The expected spot market price without the presence of futures contracts is always higher (ranges from 84 to 111\euro/MWh) than the expected spot price in the presence of futures contracts (ranges from 75 to 100\euro/MWh), which overlapped in the CFD model. In the CFD, the expected spot price without the presence of the futures market is always higher than the futures price (coincides with normal backwardation over time) and decreases with respect to RES penetration.

Therefore, buyers benefit from RES penetration as it decreases electricity prices in both stages and motivates RES generators to generate and trade higher quantities of electricity in the market. Regarding competition, there is a counter-argument that attaining a certain level of renewable with no marginal cost in a liberalized electricity market distorts market competitiveness and creates contango \citep{blazquez2018renewable}. The argument is that the further reduction of expected spot prices beyond the speed of the futures price with respect to RES penetration is due to market inefficiency created by the dispatchability and zero generation cost of RES generators. The decrease in price with RES penetration corroborates with classical economic theory, where the higher the competition (because of RES generators competitiveness increases) the larger is the fall in price even by conventional generators to compete with RES generators. In perfect competition, generators are price takers and they have no influence on the future market prices.

As far as trading quantities are concerned, conventional generators trade quantities that range from 11,414 to 12,761MWh with RES penetration. Futures market trading for conventional generators is not significantly affected as much as the expected spot market quantities, which decrease across RES penetration (see \cref{L_f1}). Following the market prices trend, the CVaR decreases for conventional generators with respect to RES penetration, which shows left tailed profit distributions with a low probability of low profits with high RES penetration. This is evident as the RES generators' trading quantities increase both in the futures market and spot market with respect to RES penetration (see \cref{L_f2}). As profit is the reflection of price and quantity, the higher the RES penetration, the lower the conventional production, which in turn lowers their profit.

Comparing the models, conventional generators' profit is higher in the GM than in the CFD, but it decreases in both models with respect to RES penetration. This is due to the higher futures electricity price in the GM than the relatively lower futures price in the CFD with the RES penetration. RES generators' profit, on the other hand, increases in both models as generators trade higher electricity with respect to RES penetration. 

When generators react in the market as perfect competition, the competition squeezes their profits compared to Cournot competition. This is an intuitive result as it shows when generators act in the market as perfect competition, the pattern of profits follows the shape of prices, both in the GM and in the CFD, which is consistent with existing literature \citep{jabr2005robust,oliveira2013contract}.

\begin{figure}
	\centering
	\includegraphics[width=\textwidth]{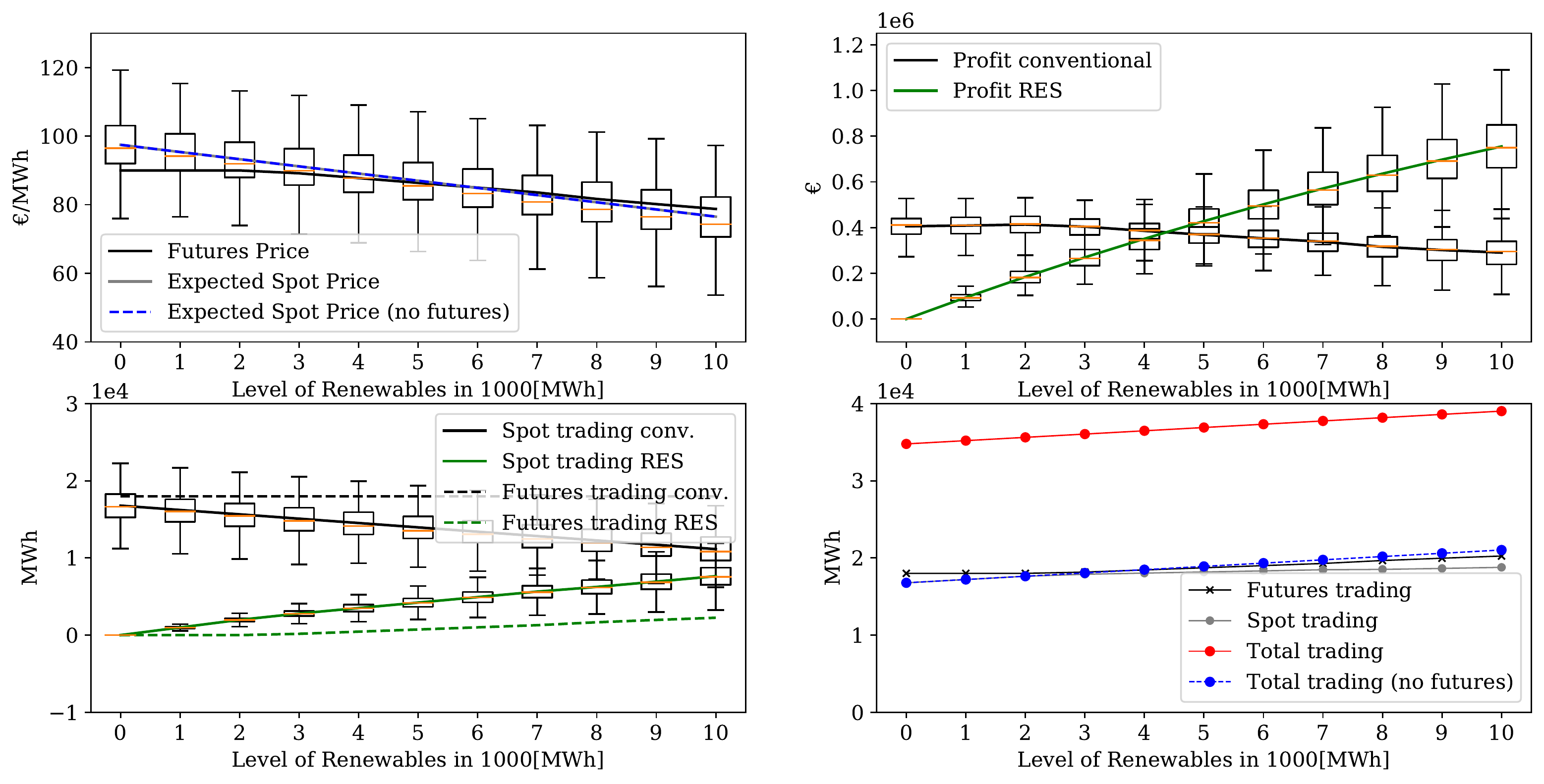}
	\caption{Risk averse perfect competition in the CFD}
	\label{L_f4}
\end{figure}

Total trading quantities in the CFD are much higher than the GM. This is because the CFD model includes physical and financial quantities (\cref{L_f2} and \cref{L_f4}). In other words, the CFD entails larger electricity transactions in the spot market, which is explained by the lack of a physical settlement in the futures market that requires all quantities must be delivered in the spot market. In addition, total trading without the presence of futures market is higher in the GM and lower in the CFD, where the expected spot market trading is higher. Conversely, the spot market total trading with the futures market decreases in the GM with high RES penetration.
\begin{figure}
	\centering
	\includegraphics[width=\textwidth]{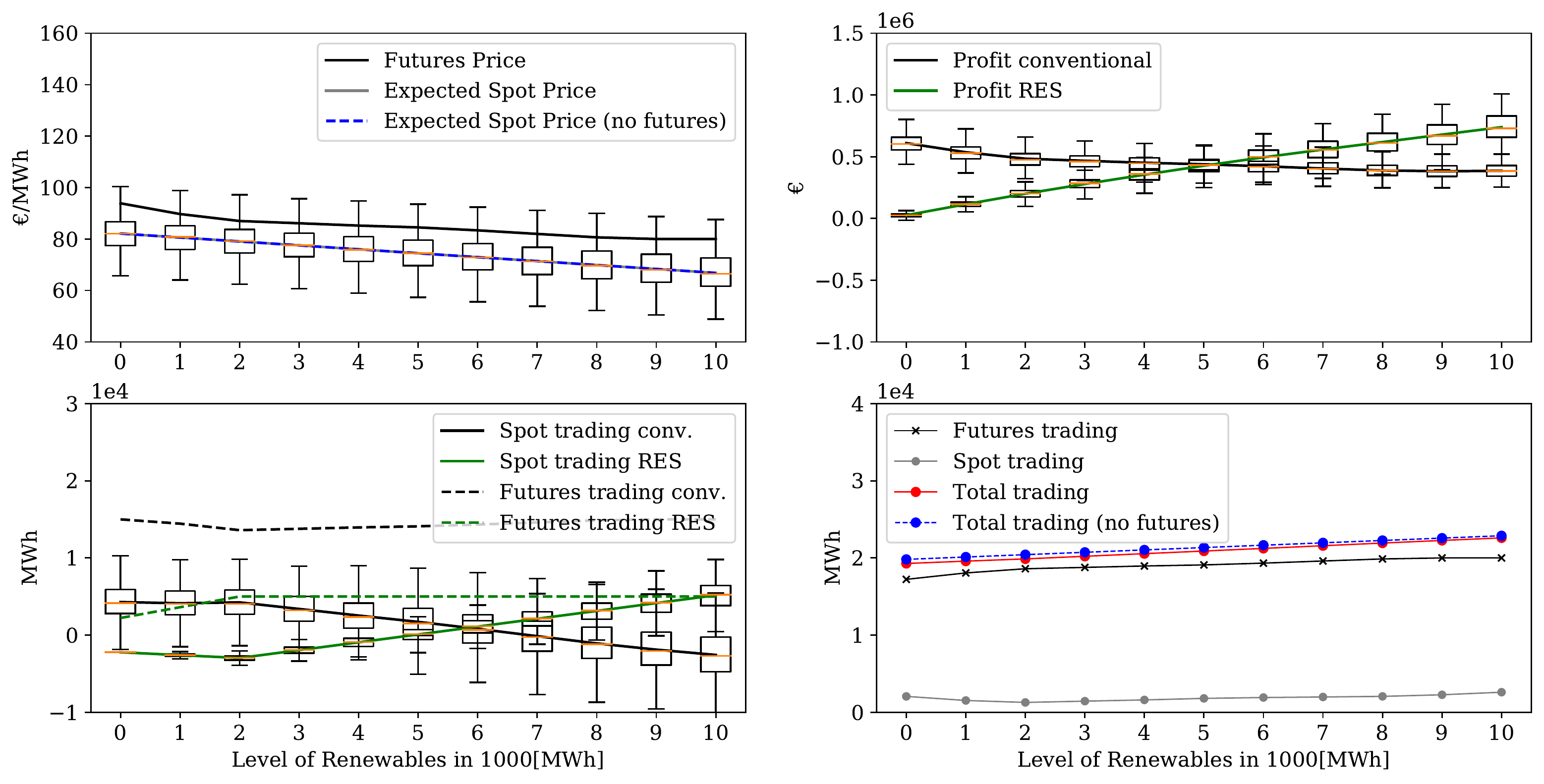}
	\caption{Risk averse perfect competition in the GM}
	\label{L_f3}
\end{figure}

\begin{table}
	\centering
	\caption{The minimum and the maximum electricity prices [\euro/MWh] with the two models and the two levels of competitions in the risk averse case.}
	\begin{tabular}{lll|ll}
		\hline\hline
		& \multicolumn{2}{l|}{General Model (GM)} & \multicolumn{2}{l}{Contract for differences (CFD)} \\ \cline{2-5} 
		Price &
		\begin{tabular}[c]{@{}l@{}}Cournot\\  model\end{tabular} &
		\begin{tabular}[c]{@{}l@{}}Perfect\\ Competition\end{tabular} &
		\begin{tabular}[c]{@{}l@{}}Cournot\\  model\end{tabular} &
		\begin{tabular}[c]{@{}l@{}}Perfect\\ Competition\end{tabular} \\ \hline
		$P^F$   & 101-116  & 83-100  & 78-90  & 79-90   \\
		$P_{\omega}^S$ & 75-100   & 76-97 & 70-96 & 76-97 \\
		$P_{\omega}^S$ only & 84-111 & 76-97 & 84-111 & 76-97 \\ \hline\hline
	\end{tabular}
	\label{tab_2}
\end{table}
\cref{L_f3} and \cref{L_f4} show the market outcomes with perfect competition in the GM and in the CFD, respectively. With perfect competition, the overall price is lower in both models with respect to RES penetration (ranges from 76 to 100\euro/MWh in both contracts). Besides, the expected spot prices are overlapped in both the GM and the CFD (with and without the presence of the futures market). 
In the GM, total trading quantity with and without the presence of futures contracts shows no significant difference as presented in \cref{L_f3}. However, this is the opposite as the CFD model encompasses physical and financial quantities. In the CFD, total trading sharply increases with no futures market despite the trading volume in the spot market with the presence of the futures market is not as low as it is in the GM.
\subsection{Risk neutral generators' numerical results (\texorpdfstring{$\phi=0$)}{Lg}}\label{ap_1}
\begin{figure}
	\centering
	\includegraphics[width=\textwidth]{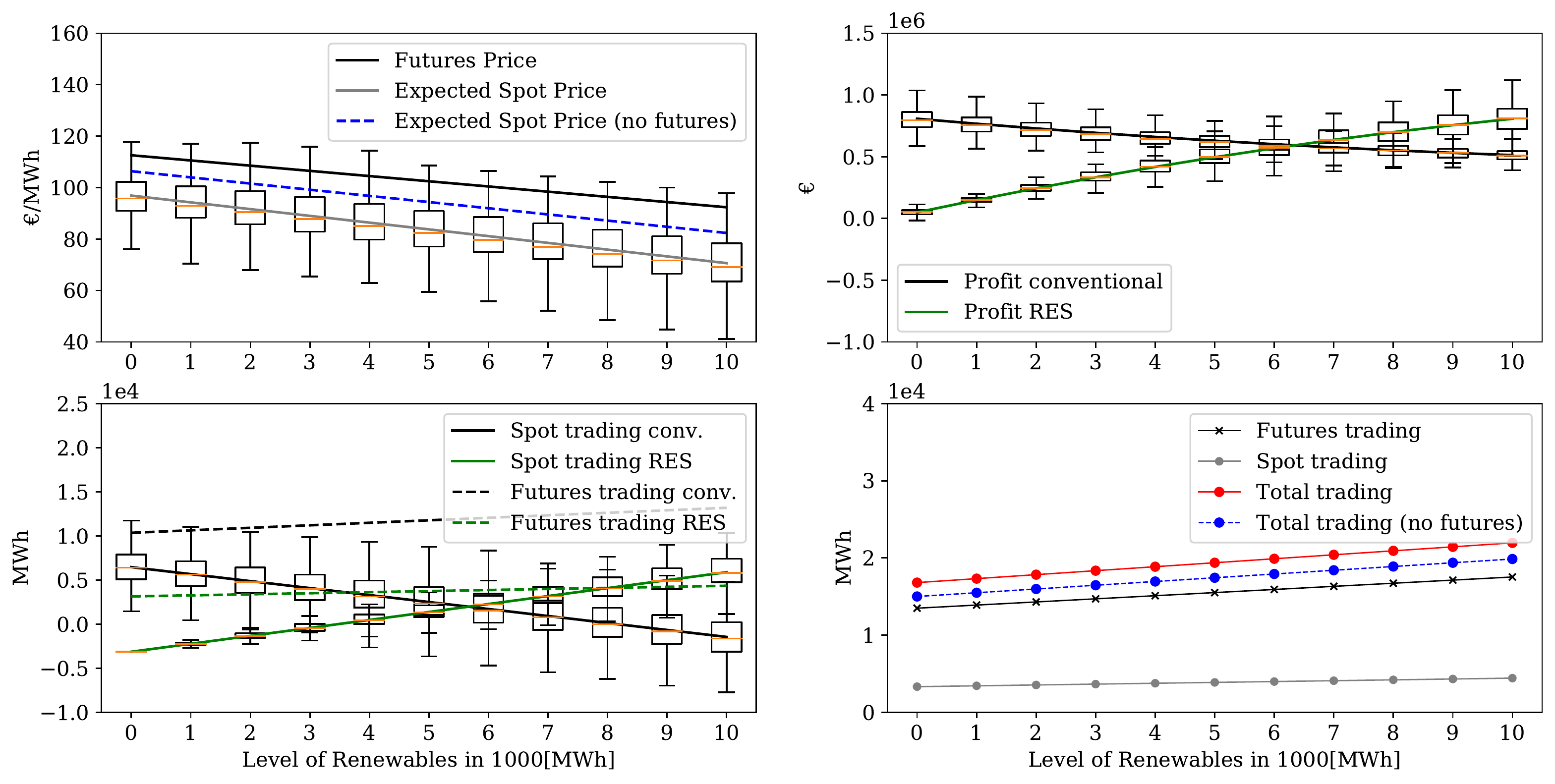}
	\caption{Risk neutral Cournot competition in the GM with RES penetration.}
	\label{L_f_1}
\end{figure}
\begin{figure}
	\centering
	\includegraphics[width=\textwidth]{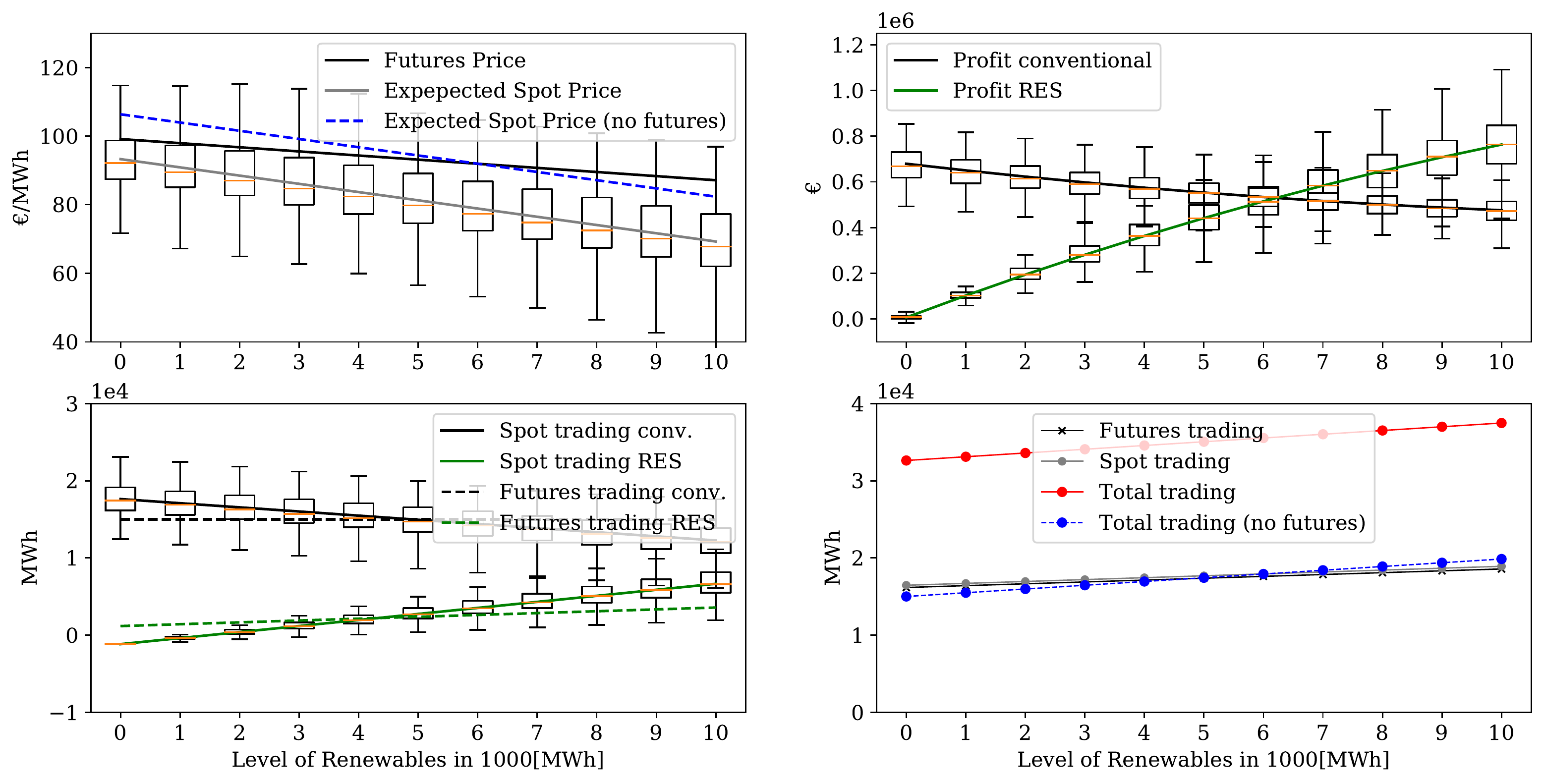}
	\caption{Risk neutral Cournot competition in the CFD with RES penetration.}
	\label{L_f_2}
\end{figure}
\begin{figure}
	\centering
	\includegraphics[width=\textwidth]{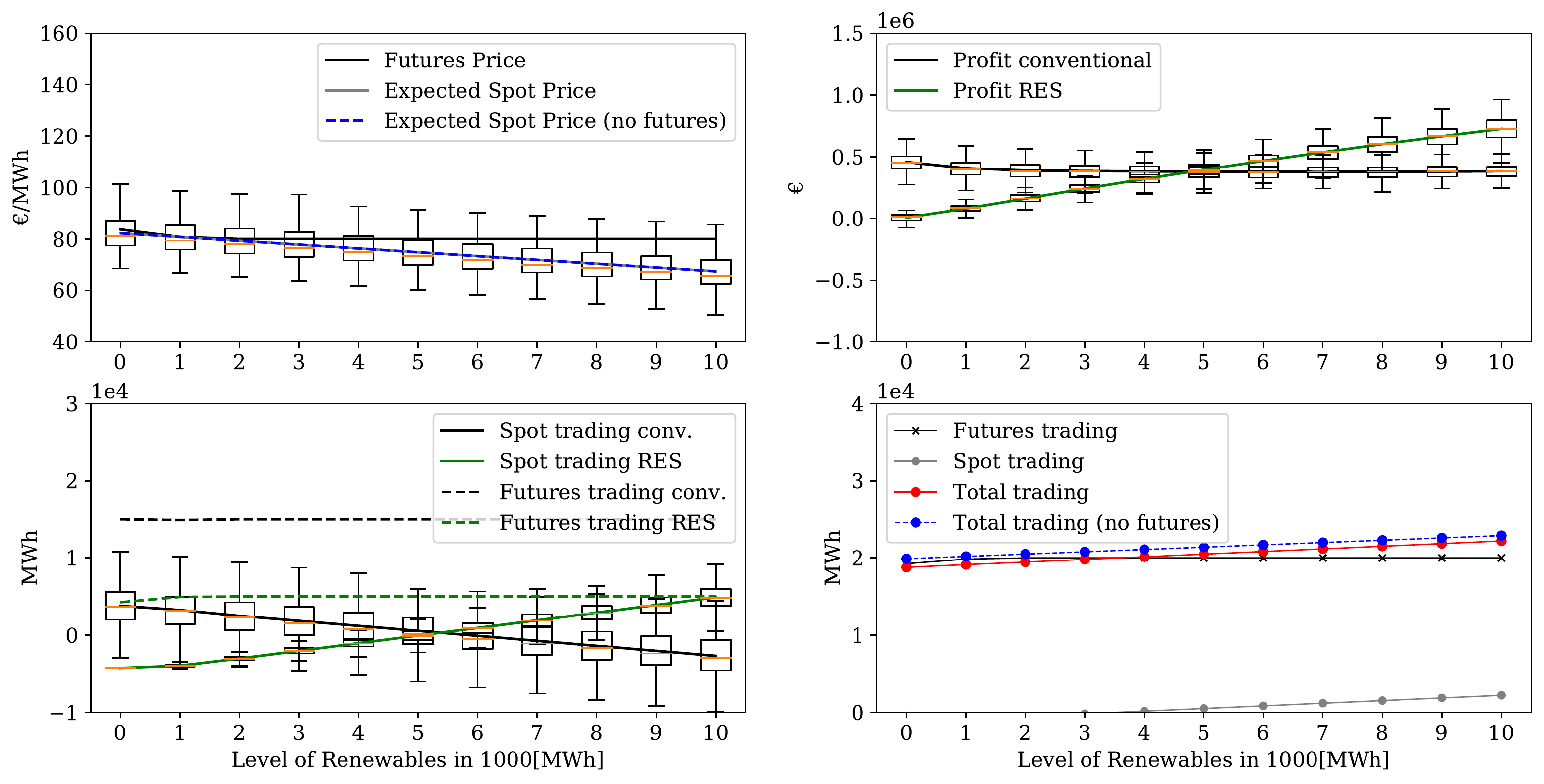}
	\caption{Risk neutral perfect competition in the GM with RES penetration.}
	\label{L_f_3}
\end{figure}
\begin{figure}
	\centering
	\includegraphics[width=\textwidth]{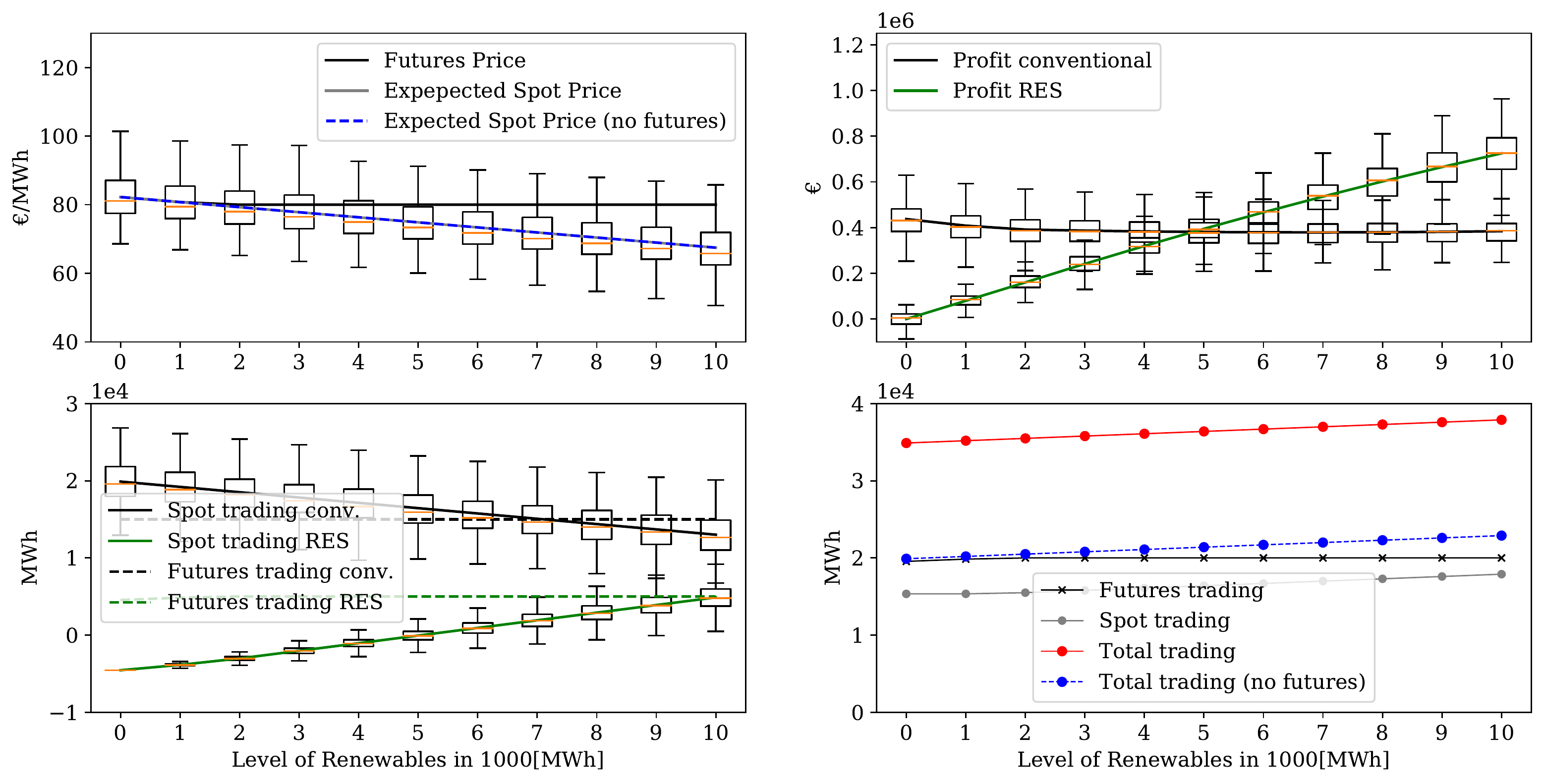}
	\caption{Risk neutral perfect competition in the CFD with RES penetration.}
	\label{L_f_4}
\end{figure}
In the risk neutral case, the numerical simulation is done by setting the risk parameter $\phi=0$. In this case, generators maximize their expected profits, which is expressed in (\ref{L_eq:5}). \cref{L_f_1} and \cref{L_f_2} show the risk neutral results for Cournot competition, and \cref{L_f_3} and \cref{L_f_4} perfect competition with respect to high RES penetration.
Similar to the risk averse case, the overall electricity price decreases with respect to RES penetration. With Cournot competition, the market is in contango in the GM, where the futures price is higher than the expected spot prices, and in normal backwardation in the CFD model, where the futures price is lower than the expected spot market price with respect to RES penetration. 

Contrary to the risk averse generators case in the GM, conventional generators’ futures market trading in the risk neutral case slightly increases with RES penetration, whereas RES generators' futures market trading is higher in the risk averse case with Cournot competition (comparing \cref{L_f1} and \cref{L_f_1}). In the CFD, conventional generators’ trade with Cournot competition is always higher than the RES generators’ trading with respect to RES penetration despite RES generator profit is as high as conventional generators’ profit in the GM.   
For the perfect competition, futures prices overlapped with expected spot market prices with respect to RES penetration, in both the GM and the CFD model, which is inline with the standard non-arbitrage condition that explains futures price is equal to the expected spot market prices, particularly with risk neutrality. Total trading and profit exhibit similar pattern with the Cournot model discussed above.
\begin{table}
	\centering
	\caption{Expected market outcomes with levels of competition in the GM and CFD for risk neutral and risk averse cases.}
	\resizebox{\textwidth}{!}{%
		\begin{tabular}{lllll|llll}
			\hline\hline
			& \multicolumn{4}{l|}{Cournot Model}         & \multicolumn{4}{l}{Perfect competition}     \\ \cline{2-9} 
			Market & \multicolumn{2}{l}{Risk neutral} & \multicolumn{2}{l|}{Risk averse} & \multicolumn{2}{l}{Risk neutral} & \multicolumn{2}{l}{Risk averse} \\ \cline{2-9} 
			Outcomes & GM       & CFD       & GM       & CFD      & GM       & CFD       & GM        & CFD      \\ \hline 
			$P^F$   & 108.28   & 108.28    & 107.68   & 107.68   & 87.26    & 87.26     & 91.46     & 91.46    \\
			$P_{i\omega}^S$ & 90.64    & 90.64     & 88.48    & 88.48    & 87.26    & 87.26     & 86.99     & 86.99    \\
			$\sum_{i\in {I}}q_i^F$ & 17999.99 & 10815.77  & 11791.94 & 17999.99 & 17999.99 & 14658.56  & 13924.75  & 17999.99 \\
			$\sum_{i\in {I}}q_i^S$ & 14737.43 & 2740.58   & 1656.61  & 14467.78 & 14106.44 & -550.57   & 19.05     & 13961.32 \\
			$\sum_{j\in {J}}q_j^F$ & 667.27   & 3527.72   & 2671.51  & 1065.89  & 900.89   & 3888.64   & 3782.41   & 863.93   \\
			$\sum_{j\in {J}}q_{j\omega}^S$ & 4321.95  & 1461.51   & 2441.14  & 4046.76  & 4251.07  & 1263.32   & 1160.34   & 4078.83  \\
			$\sum_{i\in {I}}\Pi_i$& 394779.9 & 634928.70 & 636130.7 & 362298.2 & 361638.9 & 393461.36 & 446717.43 & 362277.3 \\
			$\sum_{j\in {J}}\Pi_j$& 397829.4 & 483218.95 & 478013.2 & 401020.6 & 424595.6 & 424595.63 & 426842.08 & 408029.3\\  \hline\hline
	\end{tabular}}
	\label{tab_3}
\end{table}

Our results attest to the literature that high renewables integration is associated with exposure to electricity market risks \citep{schleicher2012renewables,klessmann2008pros}. \cref{L_f_2} shows that market outcomes are more dispersed as the RES deployment increases in the electricity market. When intermittent generation increases, it also increases the dispatching cost of conventional generators, as demand requires the combination of RES with flexible powers with high variable costs.

\subsection{Overall comparison}
\cref{tab_3} shows the market outcomes, mainly the overall expected outcomes for comparison between risk neutrality and risk aversion, levels of competition, and the two contract settlements. Futures market and expected spot market prices are slightly higher for risk neutral generators in the Cournot competition than in the perfect competition. Expected spot market trading varies in all the cases presented in \cref{tab_3}, mainly for the futures market trading which matches the expectation. RES generators trade more electricity in the spot market compared to their trade in the futures market. The sum of one large RES generator’s profit is almost as close as the three conventional generators’ profit as the RES penetration displaces conventional ones from competition.  High RES penetration results in higher standard deviations as higher levels of RES are associated with higher levels of uncertainty and exposure to higher risks.  Since RES generators are assumed with zero marginal cost (do not have cost uncertainty) in our model setting, the increase in RES generation quantity offsets the risks from demand and RES capacity uncertainties. Moreover, it is evident from our simulation results that RES generators trade much of their electricity in the spot market as contracts manage the uncertainties raised from demand and RES capacity availability. Another reason is due to non-dispatchability of RES, they have to trade all the remaining quantities in the spot market so that it may contribute to the spot market trading increments. 
\subsection{The effect of risk parameter on market outcomes}
\begin{figure}
	\centering
	\includegraphics[width=\textwidth]{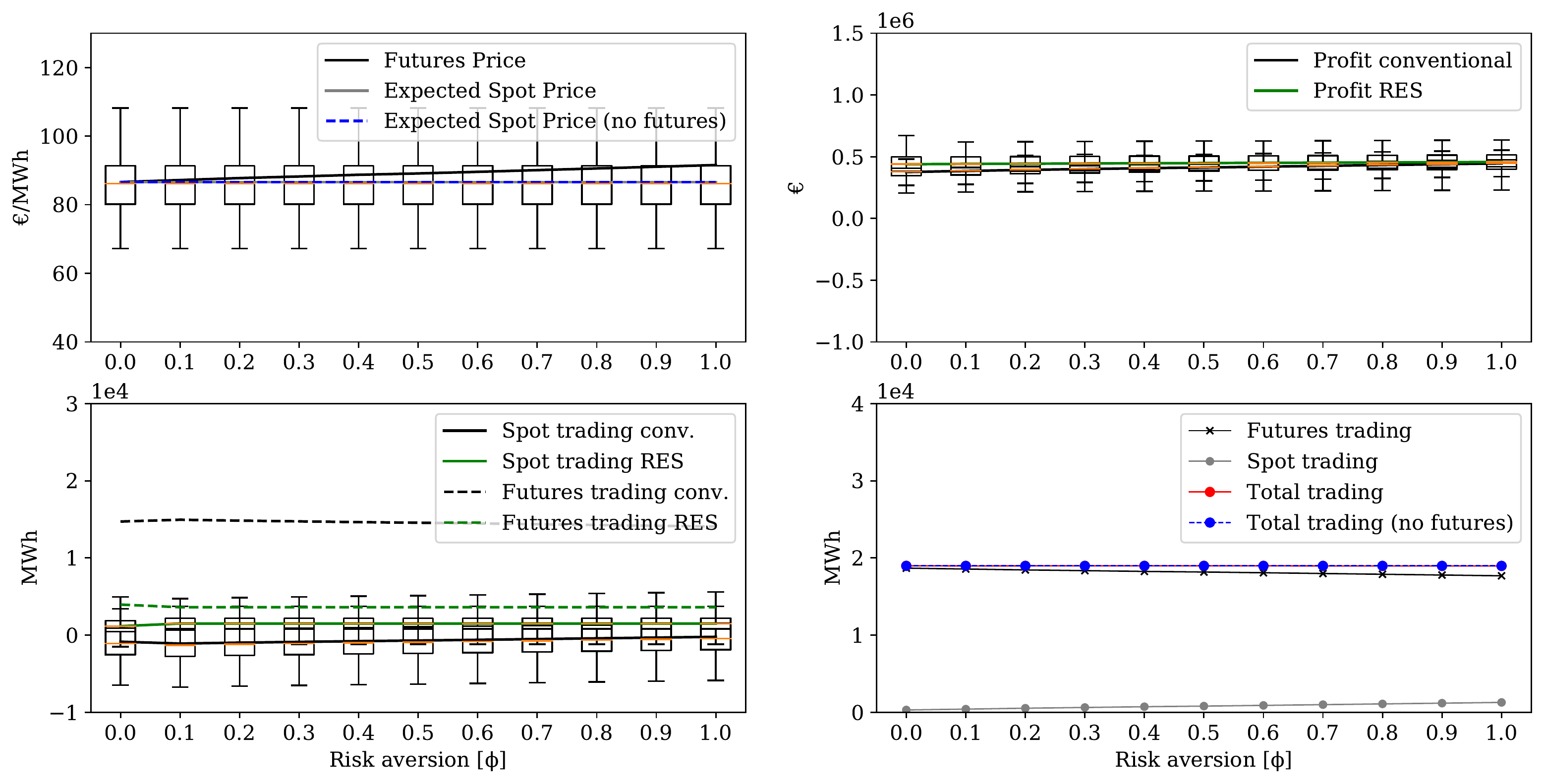}
	\caption{The effect of varying risk parameter on market outcomes in perfect competition.}
	\label{cvar_1}
\end{figure}
Finally, we analyze the impact of risk parameter ($\phi$) increase on market outcomes with perfect competition, and in the GM as a representative reference. This is done by keeping the RES parameter at its expected value (5,000MWh) and moving the CVaR parameter (from extreme risk neutrality ($\phi=0$) to extreme risk aversion ($\phi=1$)).
With this setting, \cref{cvar_1} shows that risk aversion increases electricity market prices both in the futures market and spot market, though the effect is higher in the futures market. The expected spot price overlapped both with and without the presence of the futures contracts. For conventional generators, spot market quantity slightly increases with respect to the CVaR parameter, which is reflected in their improved profit as well. This is a counter-intuitive result as risk neutrality is expected to render higher profit. This implies contracts give a shield for generators from production cost, demand, and RES capacity uncertainties risks, as the more risk averse they are the better the confident they become to produce, and retail their electricity. 

\section{Summary and conclusions}\label{L_1202}
In this paper, we propose a game-theoretical framework to model an electricity market that considers different futures market contract designs in a two-stage approach under uncertainty and high penetration of renewables. 
 
We analyze the profit-maximizing problems of risk neutral (using expected value) and risk aversion (using the CVaR) generators with different contracts (physical and financial) under different generation technologies (RES and conventional). This is done by taking conjectural variations assumption where each generator has an estimation of the impact of its production may have on market prices and rival quantities.
In this regard, generators   have the option to trade their generation first, in the futures market (stage-one), and subsequently in the spot market (stage-two) given the inherent uncertainties. Since demand, conventional generators' production cost, and RES generation availability are uncertain, a coherent risk measure with the CVaR is introduced to model risk aversion. We also introduce different types of contracts in the futures market to evaluate their performance and impact in the equilibrium market outcomes, which in turn depend on the levels of RES generation in the system.

The analytical derivation of the market equilibrium starts at the second stage (spot market) where we parameterize the spot market equilibrium as a function of the futures market decision variables. In the first stage, the global equilibrium of the market is computed from the joint solution of all the generators' profit maximization problems using the CVaR. The global optimization is solved by concatenating and minimizing the product of the complementarity constraints subject to the remaining equality and inequality KKT optimality conditions from the CVaR formulation, and the closed-form solutions obtained from the spot market equilibrium.

The analytical results obtained are tested with numerical results by analyzing different levels of competition (Cournot and perfect competition), different contracts (physical and financial) with risk neutral and risk averse strategies for players. 

We can conclude that the level of risk aversion and competition have a strong impact on electricity market outcomes with respect to RES penetration. Therefore, RES penetration increases overall quantity traded and decreases overall electricity prices, which is better for social welfare. RES generators’ profit increases as their trading quantities increase and contracts provide a way for generators to manage the risk associated with the inherent volatility of the spot market. The simulation results show that generators can increase their profit when they act as risk averse in the market, which is a counter-intuitive result of the model.
 
This may be because of the positive impact financial contracts have in risk hedging which, in turn, increases RES generators trading quantity mainly from RES capacity. Combining financial contracts with a coherent risk measure can help managers to deal with electricity markets with high uncertainty levels, while achieving the energy transition targets. The proposed approach can help decision makers to identify an appropriate modeling strategy, and to choose the best type of contracts to hedge risk in such two-stage oligopolistic markets. To this end, as a contribution to the existing literature, we provide a risk-management strategy via financial derivatives for improving the viability of energy transition, as well as policy insights regarding application to liquid futures market.

Comparing contracts, the CFD (financial trading) model renders better consumer surplus as it has lower electricity prices while the GM (physical delivery) is better for electricity generators as it renders higher electricity prices and higher profits. 

There are several possible extensions to the model and methods proposed in this paper. First, the conventional generators’ profit function could be modified to study emission trading scheme in the electricity market in a bid to plummet greenhouse gas emission. Second, by relaxing conventional generators’ production cost function to be linear, the model could be extended to study the interaction between market power, and risk aversion considering different contracts over time. Finally, the model can be applied for problems that incorporate financial contracts and risk aversion to study equilibrium solutions.

	\section*{Acknowledgements}
	The authors sincerely thank the editor and the reviewers for their constructive comments and suggestions that improve this study. Carlos Ruiz gratefully acknowledges the financial support from the Spanish government through projects PID2020-116694GB-I00 and from the Madrid Government (Comunidad de Madrid) under the Multiannual Agreement with UC3M in the line of “Fostering Young Doctors Research” (ZEROGASPAIN-CM-UC3M), and in the context of the V PRICIT (Regional Programme of Research and Technological Innovation.
	\bibliographystyle{elsarticle-num-names}
		\newpage
		\section*{References}
	\bibliography{Reference_WWP_latest}
	\appendix
	\section{Proof for Proposition 1}\label{Prop_1}
	\begin{proof}\label{thm1}
	The profit for conventional  generator $i$, in scenario $\omega$, in the spot market is: 	
	\begin{equation}\label{L_ed1}
	\Pi_{i\omega}^S=P^S_{\omega}q_{i\omega}^S-a_{i\omega}- b_{i\omega} (q_i^F+q_{i\omega}^S)-\frac{1}{2}c_{i\omega}(q_i^F+q_{i\omega}^S)^2.
	\end{equation}
	From \eqref{L_ed1}, we can derive the spot market equilibrium by computing the first-order optimality conditions for all generators simultaneously as:
	\begin{equation}\label{L_7}
	\frac{\partial\Pi_{i\omega}}{\partial q_{i\omega}^S}=0=\frac{\partial P_{\omega}^S}{\partial q_{i\omega}^S}q_{i\omega}^S+P_{\omega}^S-b_{i\omega}-c_{i\omega}(q_{i}^F+q_{i\omega}^S)=0\quad \forall i,\forall \omega
	\end{equation}	
	where $\frac{\partial P_{\omega}^S}{\partial q_{i\omega}^S}=-\beta_{\omega}^S\left(1+\sum_{i\ne k}\frac{\partial q_{k\omega}^S}{\partial q_{i\omega}^S} \right)=-\beta_{\omega}^S(1+\delta_i).$
	By substituting $\frac{\partial P_{\omega}^S}{\partial q_{i\omega}^S}=-\beta_{\omega}^S(1+\delta_i)$ into \eqref{L_7}, we can simplify and rearrange it to solve for $q_{i\omega}^S$:
	\begin{equation}\label{L_8}
		q_{i\omega}^S=\frac{1}{\beta_{\omega}^S(1+\delta_i)+c_{i\omega}}(P_{\omega}^S-b_{i\omega}-c_{i\omega}q_i^F)=
		\tau_{i\omega}(P_{\omega}^S-b_{i\omega}-c_{i\omega}q_i^F) \quad \forall i,\forall \omega.
	\end{equation}
At this level, we can substitute $P_{\omega}^S$ from \eqref{L_4} into (\ref{L_8}) to obtain the clearing price in the spot market in terms of futures market decision variables which can be rearranged as: $P_{\omega}^S=\hat{\gamma}_{\omega}^{S}-\beta^S_{\omega}\sum_{i\in {I}}q_{i\omega}^S-\beta^S_{\omega}\sum_{i\in {I}} q_{i}^F$, where we can collect price as:
	$P_{\omega}^S+\beta^S_{\omega}\sum_{i\in {I}}\tau_{i\omega}P_{\omega}^S =\hat{\gamma}_{\omega}^{S}-\beta^S_{\omega}\sum_{i\in {I}}q_i^F+\beta^S_{\omega}\sum_{i\in {I}}\tau_{i\omega}\left( b_{i\omega}+c_{i\omega}q_i^F\right).$ Finally, the equilibrium spot market price is:
	\begin{eqnarray}\label{L_edi_3}
	P_{\omega}^S=\varphi_{\omega}\left[\hat{\gamma}_{\omega}^{S}-\beta_{\omega}^S\sum_{i\in {I}}q_{i}^F+\beta_{\omega}^S\sum_{i\in {I}}\tau_{i\omega}\left( b_{i\omega}+c_{i\omega}q_i^F\right) \right]\quad \forall\omega.
	\end{eqnarray}
\end{proof}
	\section{Proof for Proposition 3}\label{Prop_2}
	\begin{proof}
From the profit function expressed in \eqref{L_cfd_1}, we can compute the spot equilibrium by deriving the first-order optimality conditions for all generators:
	\begin{equation}\label{L_1001}
	\frac{\partial\Pi_{i\omega}}{\partial q_{i\omega}^S}=-q_i^F\frac{\partial P_{\omega}^S}{\partial q_{i\omega}^S} +\frac{\partial P_{\omega}^S}{\partial q_{i\omega}^S}q_{i\omega}^S   +P_{\omega}^S-b_{i\omega}-c_{i\omega}q_{i\omega}^S=0 
	\end{equation}
	Rearranging (\ref{L_1001}) and solving for $q_{i\omega}^S$ as:
	\begin{equation}\label{L_3.20e}
		q_{i\omega}^S=\tau_{i\omega}\left(P_{\omega}^S-b_{i\omega} +  q_i^F \beta_{\omega}^S(1+\delta_i)\right). 
	\end{equation}
	Then, substituting \eqref{L_3.20e} into \eqref{L_222} and simplifying it gives:
	$P_{\omega}^{S}+\beta_{\omega}^S\sum_{i\in {I}} \tau_{i\omega}P_{\omega}^S=\hat{\gamma}_{\omega}^{S}+\beta_{\omega}^S\sum_{i\in {I}}\tau_{i\omega}b_{i\omega}-\beta_{\omega}^S\sum_{i\in {I}}q_i^F \beta_{\omega}^S(1+\delta_i)\tau_{i\omega}$, from which the optimal price at the spot market is expressed as:
	\begin{equation}\label{L_71}
	P_{\omega}^{S}=\varphi_{\omega}\left[ \hat{\gamma}_{\omega}^{S}+\beta_{\omega}^S\sum_{i\in {I}}\tau_{i\omega}b_{i\omega}-\beta_{\omega}^S\sum_{i\in {I}}q_i^F \beta_{\omega}^S(1+\delta_i)\tau_{i\omega}\right]\quad \forall \omega. 
	\end{equation}
	Finally, by substituting \eqref{L_71} into \eqref{L_3.20e}, the optimal spot quantity for generator $i$, at scenario $\omega$, is expressed as:
	\begin{equation*}
	q_{i\omega}^S=\tau_{i\omega} \left[\varphi_{\omega}[ \hat{\gamma}_{\omega}^{S}+\beta_{\omega}^S\sum_{i\in {I}}\tau_{i\omega}b_{i\omega}-\beta_{\omega}^S\sum_{i\in {I}}q_i^F \beta_{\omega}^S(1+\delta_i)\tau_{i\omega}]
	-b_{i\omega} +  q_i^F \beta_{\omega}^S(1+\delta_i)\right] \quad\forall i, \forall\omega.
		\end{equation*}
\end{proof}
	\section{Proof for Proposition 4}\label{Prop_3}
\begin{proof}
The price-demand curve is:
\begin{eqnarray}\label{L_SO_1}
P_{\omega}^S =\gamma^S_{\omega}-\beta^S_{\omega}\left(\sum_{i\in{I}} q_{i\omega}^S+\sum_{j\in{J}}Q_{j\omega}\right)\quad\forall \omega
\end{eqnarray}
which can be further simplified as:
$P_{\omega}^{S}=\hat{\gamma}_{\omega}^{S}-\beta_{\omega}^S\sum_{i\in {I}} q_{i\omega}^S.$
  
The equilibrium in the spot market is reached when all conventional generators maximize their profits simultaneously, which is expressed by the first-order optimality conditions as:
\begin{equation}\label{L_23}
\frac{\partial\Pi_{i\omega}}{\partial q_{i\omega}^S}=\frac{\partial P_{\omega}^S}{\partial q_{i\omega}^S}q_{i\omega}^S+P_{\omega}^S-b_{i\omega}-c_{i\omega}q_{i\omega}^S=0.
\end{equation}	
Simplifying \eqref{L_23}, gives $q_{i\omega}^S=\tau_{i\omega}(P_{\omega}^S-b_{i\omega})$ hence we can substitute $q_{i\omega}^S$ into (\ref{L_SO_1}), and solving for the spot price as $P_{\omega}^S\left(1+\beta^S_{\omega}\sum_{i\in {I}}\tau_{i\omega}\right) = \hat{\gamma}_{\omega}^{S}+\beta^S_{\omega}\sum_{i\in {I}}\tau_{i\omega} b_{i\omega}$ gives the equilibrium price: 

$P_{\omega}^S=\varphi_{\omega}\left[\hat{\gamma}_{\omega}^{S}+\beta_{\omega}^S\sum_{i\in {I}}\tau_{i\omega} b_{i\omega} \right]$, which can be expressed by substituting $\hat{\gamma}_{\omega}^{S}$ as:
\begin{equation}\label{L_44}
P_{\omega}^S=\varphi_{\omega}\left[\gamma_{\omega}^{S}-\beta_{\omega}^S\sum\limits_{i=1}^J Q_{j\omega}+\beta_{\omega}^S\sum_{i\in {I}}\tau_{i\omega}b_{i\omega}\right] \quad \forall\omega. 
\end{equation}
Finally, substituting \eqref{L_44} into \eqref{L_23} gives the equilibrium quantity for conventional generators as:
\begin{equation}\label{L_43}
q_{i\omega}^S=\tau_{i\omega}\varphi_{\omega}\left[\hat{\gamma}_{\omega}^{S}+\beta_{\omega}^S\sum_{i\in {I}}\tau_{i\omega}b_{i\omega}\right]-\tau_{i\omega}b_{i\omega}\quad \forall i,\forall\omega. 
\end{equation}
\end{proof}
\end{document}